\documentclass[oldversion]{aa}
\usepackage{graphicx}

\usepackage{amsmath}
\usepackage{natbib}
\usepackage{epsfig}
\usepackage{txfonts}

\newcommand{\HeIlevel}[4]{{#1^{#2} {\rm #3}_{#4}}}   
\newcommand{\nuc}{{\nu_{\rm c}}}
\newcommand{\zem}{{z_{\rm em}}}
\newcommand{\ye}{{y_{\rm e}}}

\newcommand{\xg}{{x_{\gamma}}}
\newcommand{\xgi}{{x_{\gamma, 0}}}
\newcommand{\Ne}{{N_{\rm e}}}
\newcommand{\sigT}{{\sigma_{\rm T}}}
\newcommand{\me}{{m_{\rm e}}}
\newcommand{\mHe}{{m_{\rm He}}}
\newcommand{\Te}{{T_{\rm e}}}
\newcommand{\Tg}{{T_{\gamma}}}
\newcommand{\nucH}{{\nu^{\rm H}_{\rm c}}}
\newcommand{\tauS}{{\tau_{\rm S}}}
\newcommand{\tauL}{{\tau_{\rm L}}}
\newcommand{\tauc}{{\tau_{\rm c}}}
\newcommand{\tauct}{{\tilde{\tau}_{\rm c}}}
\newcommand{\etac}{{\eta_{\rm c}}}
\newcommand{\DeltanuD}{{\Delta\nu_{\rm D}}}
\newcommand{\nmax}{n_{\rm max}}

\newcommand{\pot}[2]{#1 \times 10^{#2}}

\newcommand{\HI}{{\ion{H}{i} }}
\newcommand{\HeI}{{\ion{He}{i} }}
\newcommand{\HeII}{{\ion{He}{ii} }}
\newcommand{\HeILya}{{\HeI
$\HeIlevel{2}{1}{P}{1}-\HeIlevel{1}{1}{S}{0}$ }}
\newcommand{\HeIInt}{{\HeI
$\HeIlevel{2}{3}{P}{1}-\HeIlevel{1}{1}{S}{0}$ intercombination}}
\newcommand{\HeIILya}{{\HeII Lyman-$\alpha$ }}
\newcommand{\beal}{\begin{align}}
\newcommand{\id}{{\,\rm d}}
\newcommand{\bsub}{\begin{subequations}}
\newcommand{\esub}{\end{subequations}}

\begin{document}

\titlerunning{Cosmological Helium Recombination} 
\title{Lines in the cosmic microwave background spectrum from the 
epoch of cosmological helium recombination}

\author{J. A. Rubi\~no-Mart\'{\i}n \inst{1}
          \and
          J. Chluba\inst{2} 
          \and 
          R. A. Sunyaev\inst{2,3} 
}

\authorrunning{J.A. Rubi\~no-Mart\'{\i}n et al.}

\institute{Instituto de Astrofisica de Canarias (IAC),
       C/Via Lactea, s/n, E-38200, La Laguna, Tenerife (Spain)
         \and
         Max-Planck-Institut f\"ur Astrophysik, Karl-Schwarzschild-Str. 1,
         85741 Garching bei M\"unchen, Germany
         \and 
         Space Research Institute, Russian Academy of Sciences, Profsoyuznaya 84/32,
         117997 Moscow, Russia
             }

\offprints{J. A. Rubi\~no-Mart\'{\i}n or J. Chluba, 
\\ \email{jose.alberto.rubino@iac.es} 
\\ \email{jchluba@mpa-garching.mpg.de} 
}

\date{Received / Accepted}




\abstract{The main goal of this work is to calculate the contributions to the
  cosmological recombination spectrum due to bound-bound transitions of
  helium.
  We show that due to the presence of helium in the early Universe unique
  features appear in the total cosmological recombination spectrum. These may
  provide a unique observational possibility to determine the relative abundance
  of primordial helium, well before the formation of first stars.
  We include the effect of the tiny fraction of neutral hydrogen atoms on the
  dynamics of $\ion{He}{ii}\rightarrow \ion{He}{i}$ recombination at redshifts
  $z\sim 2500$. As discussed recently, this process significantly accelerates
  $\ion{He}{ii}\rightarrow \ion{He}{i}$ recombination, resulting in rather
  narrow and distinct features in the associated recombination spectrum.
  In addition this process induces some emission within the hydrogen
  Lyman-$\alpha$ line, before the actual epoch of hydrogen recombination round
  $z\sim 1100-1500$.
  We also show that some of the fine structure transitions of neutral helium
  appear in absorption, again leaving unique traces in the Cosmic Microwave
  Background blackbody spectrum, which may allow to confirm our understanding of
  the early Universe and detailed atomic physics.
}

   \keywords{atomic processes -- cosmic microwave background -- cosmology: theory -- early Universe }

   \maketitle
%

%
\section{Introduction}
The recombination of helium practically does not influence the Cosmic
Microwave Background (CMB) angular fluctuations, as measured with great success
by {\sc Wmap} \citep{WMAP_params}, since it occurred well before the Thomson
visibility function defined by hydrogen recombination \citep{Sunyaev1970}
reaches its maximum.
However, similar to the release of photons during the epoch of cosmological
    hydrogen recombination \citep{Jose2006, Chluba2007, Chluba2006b}, one does
    expect some emission of photons by helium, and the main goal of this paper
    is to calculate the contributions to the cosmological recombination spectrum
    due to bound-bound transitions of helium.

In our recent papers we computed the detailed cosmological recombination
spectrum of hydrogen resulting from bound-bound \citep{Jose2006, Chluba2007} and
bound-free \citep{Chluba2006b} transitions between atomic levels, including up
to 100 shells, also taking into account the evolution of individual
energetically degenerate angular momentum sub-states.
We followed the ideas and suggestions of earlier investigations
\citep{Zeldovich68, Peebles68, Dubrovich1975, Bernshtein1977, Beigman1978,
  RybickiDell1993, Dubrovich95, Burgin2003, Dubrovich2004, Kholu2005,
  Wong2006}.
Observations of these recombinational lines might provide an additional
unbiased way to {\it directly} determine the baryon density of the
Universe (e.g.  see \citet{Dubrovich1975} and \citet{Bernshtein1977}, or more
recently \citet{Kholu2005} and \citet{Chluba2007d}) and to obtain some
additional information about the other key cosmological parameters, facing
different degeneracies and observational challenges.

Obviously, direct evidence for the emission of extra $\sim 5$ photons per
recombining hydrogen atom \citep{Chluba2006b} will be an {\it unique proof}
for the completeness of our understanding of the processes occurring at
redshifts $z\sim 1400$, i.e. before the CMB angular fluctuations were actually
formed.
From this point of view an observation of lines emitted during
$\ion{He}{iii}\rightarrow \ion{He}{ii}$ close to $z\sim 6000$, and
$\ion{He}{ii}\rightarrow \ion{He}{i}$ around $z\sim 2500$ will be an even more
impressive confirmation of the predictions within the standard hot big bang
model of the Universe, realising that nowadays exact computations using the
full strength of atomic physics, kinetics and radiative transfer in principle
should allow a prediction of the cosmological recombination spectrum from both
epochs with very high precision.

The first attempt to estimate the emission arising from helium recombination
was made by \citet{DubroStol1997}. However, only now detailed numerical
computations are becoming feasible, also due to the fact that atomic
physicists began to publish {\it accurate} and {\it user-friendly} transition
rates \citep{DrakeMorton2007, PCBeigman2007} for neutral helium, including
singlet-triplet transitions, which very strongly influence the recombination
of helium.

\begin{figure*}
  \centering
  \resizebox{0.95\hsize}{!}{\includegraphics[angle=90]{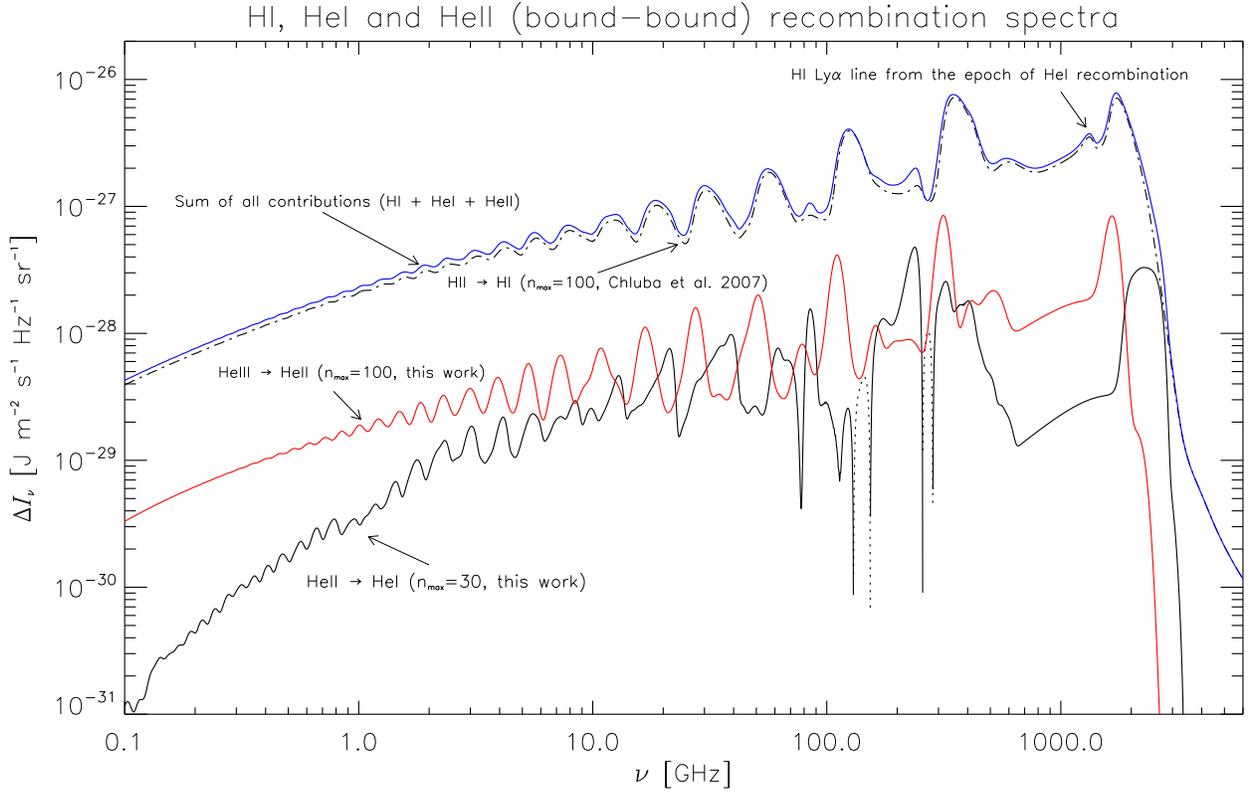}}
  \caption{Full helium and hydrogen (bound-bound) recombination spectra. The
  following cases are shown: (a) the $\ion{He}{ii}\rightarrow \ion{He}{i}$
  recombination spectrum (black solid line), which has been obtained including
  up to $\nmax=30$ shells, and considering all the J-resolved transitions up to
  $n=10$. In this case, there are two negative features, which are shown (in
  absolute value) as dotted lines; (b) the
  $\ion{He}{iii}\rightarrow\ion{He}{ii}$ recombination spectrum (red solid
  line), where we include $\nmax=100$ shells, resolving all the angular momentum
  sub-levels and including the effect of Doppler broadening due to scattering
  off free electrons; (c) the \ion{H}{i} recombination spectrum, where we plot
  the result from \citet{Chluba2007} up to $\nmax=100$.
The \ion{H}{i} Lyman-$\alpha$ line arising in the epoch of \ion{He}{i}
  recombination is also added to the hydrogen spectrum (see the feature around
  $\nu=1300$~GHz). In all three cases, the two-photon decay continuum of the
  $n=2$ shell was also incorporated. 
Feedback processes for the \ion{He}{ii} and \ion{He}{i} recombinations are not
  taken into account. Blue line shows the total recombination spectrum.  }
\label{fig:final}%
\end{figure*}

According to the computations of nuclear reactions in the early Universe
\citep{Olive1995, Cyburt2004}, the abundance of helium is close to $8\%$
percent of the number of hydrogen atoms, so naively only small additional
distortions of the CMB blackbody spectrum due to helium recombination are
expected.
However, for helium there are {\it two} epochs of recombination, a fact that
at least doubles the possible amount of additional photons.
Furthermore, $\ion{He}{iii}\rightarrow\ion{He}{ii}$ recombination is very
fast, in particular because there is a large quasi-constant amount of free
electrons belonging to hydrogen \citep{DubroStol1997}. This implies that
photons are emitted in a much shorter period, so that more narrow features are
produced\footnote{As we will demonstrate here, even the scattering of photons
  by free electrons cannot change this conclusion (see
  Sect.~\ref{sec:spec.e-scatt}).}.
It is also very impressive that the $\ion{He}{iii}\rightarrow\ion{He}{ii}$
recombination lines practically coincide and therefore amplify the
corresponding hydrogen line (see Fig.~\ref{fig:final}).
This is because the difference in the redshifts of the two recombinations is
close $\sim 4.3$, when on the other hand the energy of similar transitions
scales as $Z^2=4$ for $\ion{He}{ii}$, such that the two effects practically
compensate eachother.

The spectral distortion due to $\ion{He}{ii}\rightarrow\ion{He}{i}$
recombination should have a completely different character. First, for small
$n$ neutral helium has a much more complicated spectrum than hydrogenic atoms
(e.g.  highly probable fine-structure transitions).
In addition, the ratio of the energies for the second and first shell is $\sim
2.1$ times higher than for hydrogenic atoms, while the energies of the highly
excited levels are very close to hydrogenic.
Since the transitions from the second to the first shell are controlling
helium recombination, this leads to the situation that even for transitions
among highly excited levels the corresponding $\Delta n=1$-lines do not
coincide with those emitted during hydrogen or
$\ion{He}{iii}\rightarrow\ion{He}{ii}$ recombination.

Also it will be shown below (Sect.~\ref{sec:He.spectra}), that in the
recombinational spectrum some fine-structure lines become very bright and
that two of them are actually appearing in {\it absorption}.
These features lead to additional non-uniformities in the spectral variability
structure of the {\it total} CMB spectral distortion from recombination, where
some of the maxima are amplified and others are diminished.
This may open an unique possibility to separate the contributions of helium
and hydrogen, thereby {\it allowing to measure the pre-stellar abundance of
helium in the Universe}. Until now not even one {\it direct} method for such a
measurement is known.

For the computations of the recombinational helium spectrum we are crucially
dependent in the recombination history of helium and additional processes that
affect the standard picture strongly.
In this context, probably the most important physical mechanism is connected
with the continuum absorption of the permitted 584$\,\AA$ and intercombinational
591$\,\AA$ line by a very small amount of neutral hydrogen present in
ionizational equilibrium during the time of $\ion{He}{ii}\rightarrow\ion{He}{i}$
recombination \citep{Hu1995, Switzer2007I, Kholupenko2007}.
\citet{Switzer2007I} and \citet{Kholupenko2007} recently made the first detailed
analysis of this problem, and included it for the computations of the
$\ion{He}{ii}\rightarrow\ion{He}{i}$ recombination history, showing that the
recombination of neutral helium is significantly faster.
Here we reanalyse this process, and discuss in detail some physical aspects of
the escape problem in the aforementioned lines.

We first consider two ``extreme'' cases for the escape problem, which
can be treated analytically:
(i) where line scattering leads to {\it complete redistribution} of photons
over the line profile,
and (ii) where there is {\it no redistribution}\footnote{In this case line
scattering is totally coherent in the lab frame. During recombination this is a
very good approximation in the very distant wings of the line.}.
Moreover, we develop an useful 1D integral approximation for the escape
probability which permit us to treat any of these two cases without increasing
the computation time significantly.
Our final results for the escape probability in the more realistic case of {\it
partial redistribution} (or equivalently for coherent scattering in the rest
frame of the atom) are based on detailed numerical computations which will be
presented in a separate paper (Chluba et al. 2007, in preparation).
One can then obtain a sufficiently accurate description of the real dynamics by
{\it fudging} the escape probability using the ``no redistribution'' case
mentioned above with a certain function that can be obtained by comparison with
the full numerical computations.
Our final results for the escape probability are in very good agreement with
those obtained by \citet{Switzer2007I} (see discussion in
Sect.~\ref{sec:HIopacity}).

%
Interestingly, the hydrogen continuum process leaves additional distinct trace
in the cosmological recombination spectrum, because the continuum absorption of
the \HeI photons leads to significant early emission in the \ion{H}{i}
Ly-$\alpha$ transition at $\nu\sim 1300\,$GHz, well separated from the
Ly-$\alpha$ line originating during hydrogen recombination at $\nu\gtrsim
1500\,$GHz, and containing about $7\%$ of all photons that were released in the
hydrogen 2p-1s transition.
The amplitude and width of this feature is completely determined by the
conditions under which the above process occurs.

The main result of this paper are the bound-bound spectra of \ion{He}{ii} and
\ion{He}{i} from the epoch of cosmological recombination (see
Fig.~\ref{fig:final}).
The strongest additions to the cosmological hydrogen recombination spectrum due
to the presence of helium lines reach values up to 30-40\% in several frequency
bands. This strongly exceeds (roughly by a factor of four) the relative
abundance ratio of helium to hydrogen, raising hopes that these distortions will
be found once the recombinational lines will become observable.

It is important to note that for the computations in this paper (see
Fig.~\ref{fig:final}), we do not include the impact of feedback processes on the
computed recombinational lines. Among all the possible feedback mechanisms, the
most relevant for the recombinational spectrum is the pure continuum absorption
(far away from the resonances) of the remaining \HeIInt-line and \HeILya
photons. Due to this process, these photons will be finally absorbed, and the
corresponding features on the final spectrum will disappear, producing
additional photons that will emerge mainly through the Ly$\alpha$ line in the
hydrogen spectrum.

%
\section{Basic equations. The Helium atom}
A description of the basic formalism and equations to calculate the
time-evolution of the populations for different atomic species (hydrogen or
helium) within a multi-level code during the epoch of cosmological
recombination ($800\lesssim z \lesssim 7000$) can be found in
\citet{Seager2000}.
In this paper, we follow the same approach and notation that was used
in our previous works for the computation of the hydrogen recombination
spectrum \citep{Jose2006,Chluba2007}.
The codes used for the computations presented here were obtained as an extension
of the existing ones, by including the equations for the population of the
$\ion{He}{i}$ levels. As in our previous works, we developed two independent
implementations in order to double-check all our results.

For all the results presented in this paper we use the same values of the
  cosmological parameters which were adopted in our previous works, namely
  \citep{WMAP_params}: $\Omega_{\rm b} = 0.0444$, $\Omega_{\rm tot}=1$,
  $\Omega_{\rm m}=0.2678$, $\Omega_{\Lambda}=0.7322$, $Y_{\rm p}=0.24$ and
  $h=0.71$.

\subsection{$\ion{He}{i}$ model atom}
\label{sec:HeI}
In our computations we follow in detail the evolution of the level populations
within neutral helium, including up to $\nmax=30$ shells.
For all levels, we distinguish between ``singlet'' ($S=0$) and ``triplet''
($S=1$) states.
Up to $n=10$, we follow separately all levels with different total angular
momentum $J$. This permits us to investigate in detail the fine structure lines
appearing from cosmological recombination.
Above $n=10$, we do not resolve in $J$ quantum number, and only LS-coupling is
considered.
Each individual level is quoted using the standard ``term symbols'' as
$n^{2S+1}L_J$, and the spectroscopic notation is used.
When considering J-resolved levels, the degeneracy factor is given by $g_i =
(2J+1)$, while in opposite case we would have $g_i = (2S+1)(2L+1)$.

To completely define our model atom, we need to specify for each level
$i=\{n,L,S,J\}$, the energy, $E_i$, and photoionization cross-section,
$\sigma_{ic}(\nu)$, as a function of frequency, which is important in order to
take into account the effect of stimulated recombination to high
levels. Finally, a table with the Einstein coefficients, $A_{i\rightarrow j}$,
and the corresponding wavelengths for all the allowed transitions has to be
given.

\subsubsection{Energies}

The energies for the different levels up to $n=10$ are taken from
\cite{DrakeMorton2007}. However, this table is not absolutely complete, and some
of the high L sub-states for outer shells are missing.
In order to fill this table up to $\nmax = 30$, we proceed as follows.  We use
the formulae for quantum defects \citep[see][Chap.~11]{Drake-Book} to compute
the energies of all terms with $L\le 6$ and $n>10$.  For all other levels, we
adopt hydrogenic values for the energies, using $-R_{\rm H}/n^2$, with $R_{\rm
H}\approx 13.6\,$eV. Note that in this last case, the energy levels will be
degenerate in $L$ and $J$. However, this approximation is known to produce
very good results \citep{PCBeigman2007}.
A summary of the final energies adopted for each particular level in our
$\ion{He}{i}$ model atom is shown in Figure~\ref{fig:our_model}, both
for the singlet and triplet states.

\subsubsection{Photoionization cross-sections}

For $n<10$, the photoionization cross-sections, $\sigma_{ic}(\nu)$, are taken
from TOPbase\footnote{At http://vizier.u-strasbg.fr/topbase/topbase.html}
database \citep{topbase}. We note that this database does not contain
$J$-resolved information and only cross-sections for $L\leq 3$ can be
found. To obtain $J$-resolved cross-section we assume that the cross-section
for each sub-level is identical, i.e.
\[
\sigma_{ \{n,L,S,J\}\rightarrow c}(\nu) = \sigma_{\{n,L,S\}\rightarrow c} (\nu).
\]

There are two important issues that we would like to stress. First of all,
there are large gaps in the tables from TOPbase. Up to $n=5$, we
computed the missing cross-sections using the expressions from
\citet{Smits1996} and \citet{Smits1999} for the spontaneous photorecombination
rates to infer the photoionization rates $R_{ic}$. With this procedure it is
not possible to include the effect of stimulated recombination
self-consistently. We estimated the errors due to this effect using the
replacement $R_{ic}\rightarrow R_{ic}\times[1+n_{\rm bb}(\nuc)]$, where
$n_{\rm bb}(\nuc)$ is the blackbody photon occupation number at the ionization
threshold of the level, and found changes of the order of $10\%-20\%$.
For all other levels we follow \cite{Bauman2005} and adopt re-scaled
hydrogenic cross-section. This is done using our computations of
$\sigma_{ic}(\nu)$ for the hydrogen atom \citep[based
on][]{Karzas1961,StoreyHum1991}, and shifting the threshold frequency
accordingly.

\begin{figure*}
  \centering
  \resizebox{\columnwidth}{!}{\includegraphics{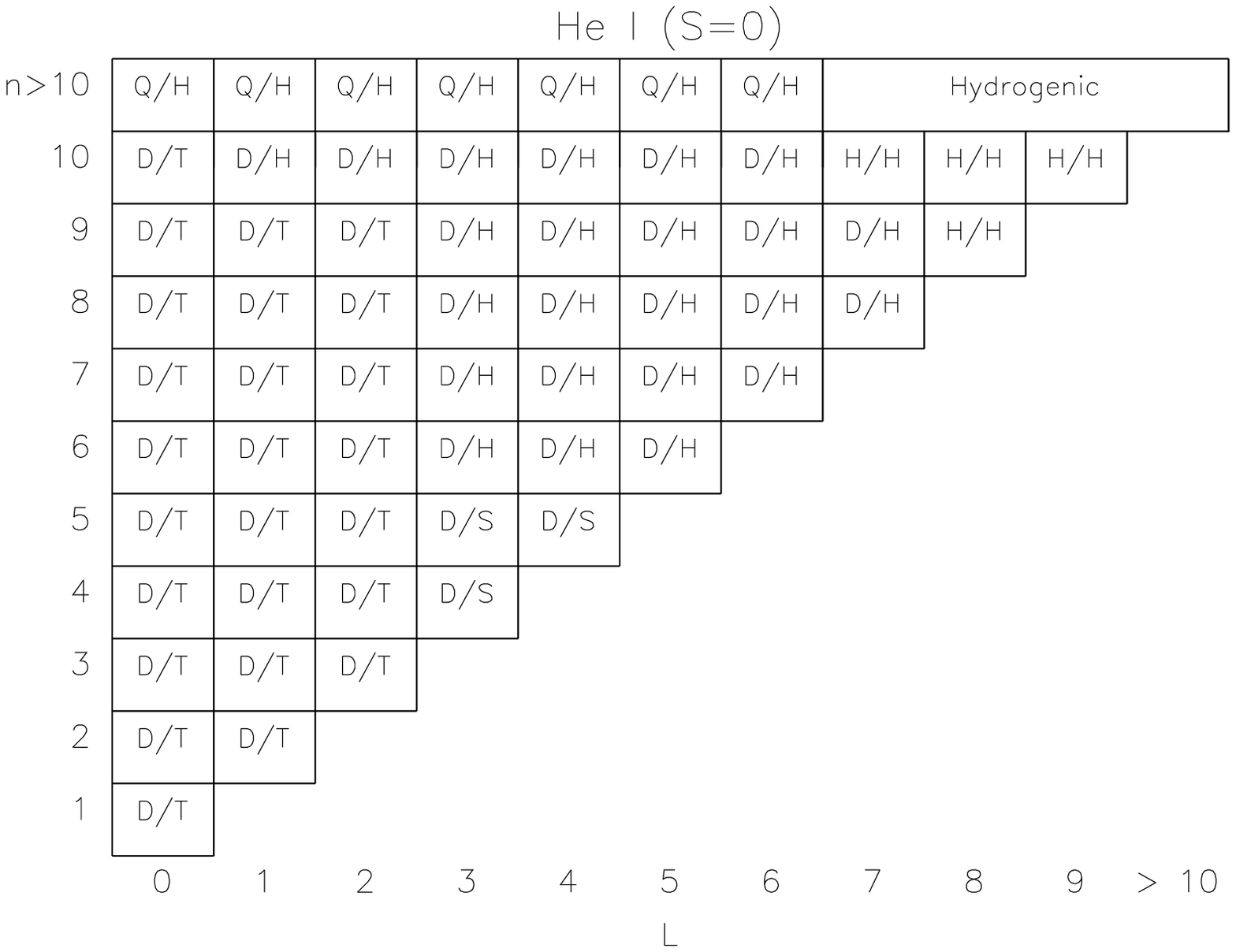}}%
  \resizebox{\columnwidth}{!}{\includegraphics{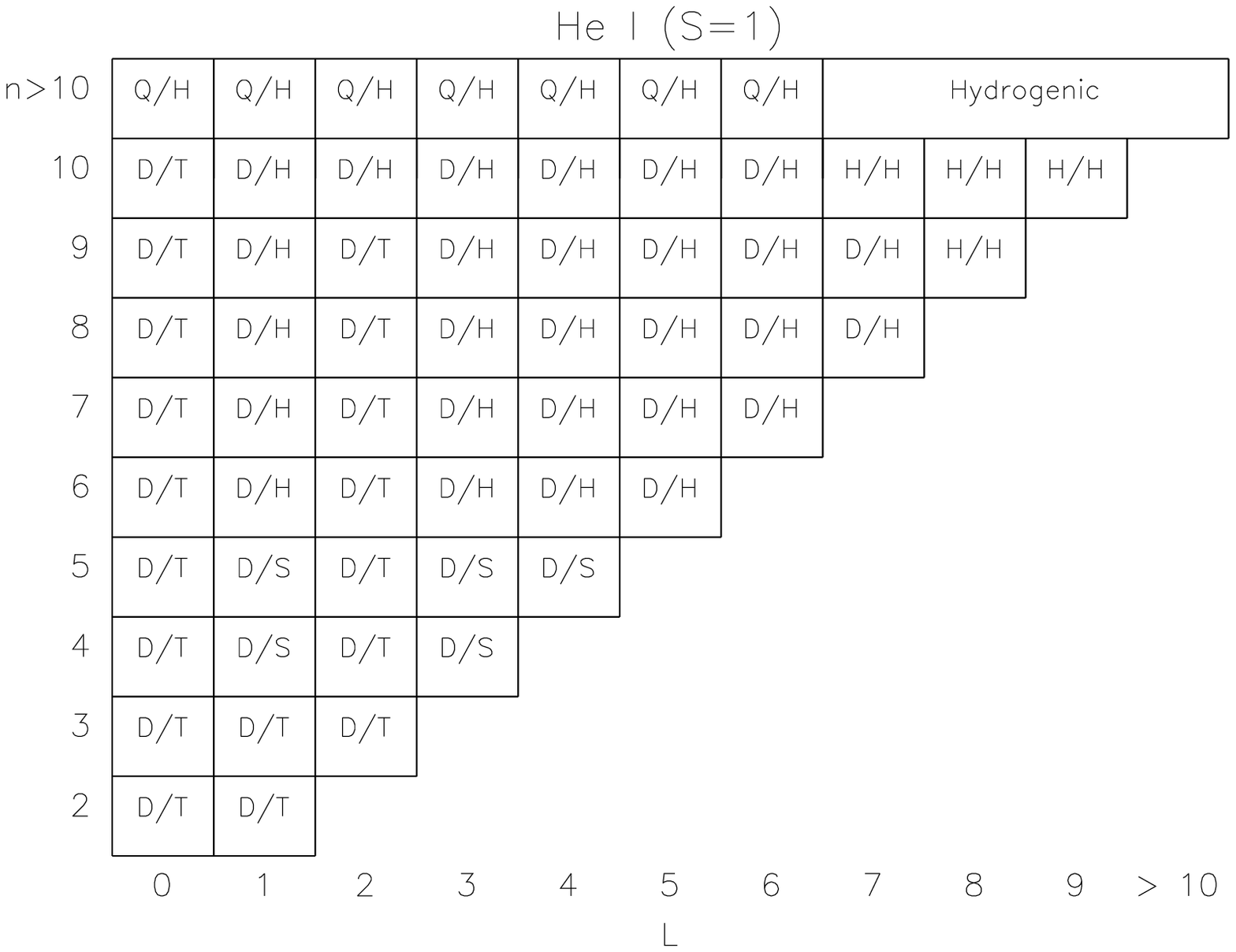}} \caption{
  Graphical representation of the values adopted for the energies and
  photoionization cross-sections of our \ion{He}{i} model. For each level, two letters
  are given, which refer to the energy and the photoionization cross section
  of that level, respectively. For the energies, the letters refer to: ``D'':
  \citet{DrakeMorton2007}; ``Q'': quantum defect expansions \cite{Drake-Book};
  and ``H'': hydrogenic approximation. For the photoionization cross-sections,
  the letters refer to: ``T'': TOPbase; ``S'': \citet{Smits1996,Smits1999};
  and ``H'': re-scaled hydrogenic values. } \label{fig:our_model}
\end{figure*}
%
%
Secondly, we would like to point out that the cross-sections provided in TOPbase
are sparsely sampled.
For example the power-law behaviour up to twice the threshold frequency is
usually given by $\sim 10$ points. Furthermore, due to auto-ionization several
resonances exist at large distances above the ionization threshold, and many of
these extremely narrow features are represented by 1 point. Fortunately, because
of the exponential cut-off from the blackbody spectrum these resonances do not
affect $R_{ic}$ significantly. Still we estimate the error budget using these
cross-section to be $\sim 10\%$.

A summary of the final values adopted for each particular level in our
$\ion{He}{i}$ model atom is also shown in Figure~\ref{fig:our_model}.
We finally note that, in order to speed up our computations, we tabulate the
photoionization rate during the initialization state of our codes, which
involves one-dimensional integral over the (blackbody) ambient photon field, and
we interpolate over this function when needed.  At every particular redshift,
the corresponding photorecombination rate is computed using the detailed balance
relation, which is satisfied with high precision due to the fact that at the
redshifts of interest, the electron temperature and the radiation temperature
practically do no differ.

\subsubsection{Transition probabilities}

Our basic database for the transition probabilities is taken from
\cite{DrakeMorton2007}, which is practically complete for the first 10 shells,
and includes 937 transitions between J-resolved states.  This database does also
contains spin-forbidden transitions (i.e. singlet-triplet and triplet-singlet),
which take into account the mixing of singlet and triplet wave functions.

However, there are some transitions missing in these tables, which involve lower
levels with $n\ge 8$ and $L\ge 7$ for the singlet, and $n>7$ and $L>6$ for the
triplet states.
These gaps are filled using re-scaled hydrogenic values as follows: for a given
transition $\{n,L,S,J\} \rightarrow \{n',L',S',J'\}$, we first obtain the non
$J$-resolved transition probability $A^{\rm He}_{nL \rightarrow n'L'}$ scaling
by the ratio of the transition frequencies to the third power. Whenever
$J$-resolved information for the level energies is available, we also compute
the weighted mean transition frequency. However, the corresponding corrections
are small. To obtain the final estimate for the $J$-resolved value, we assume
that
\begin{equation}
A^{\rm He}_{nLJ \rightarrow n'L'J'} = \frac{(2J'+1)}{(2L'+1)(2S'+1)} 
A^{\rm He}_{nL \rightarrow n'L'}
\end{equation}
These expressions are also used to include all transitions involving levels
with $n, n'>10$, and for those between $n>10$ and $n'\le 10$ states,
adopting the corresponding average over $J$ and $J'$.
For the case of $\nmax=30$, our final model contains 80,297 bound-bound
transitions.

Finally, we note that once the energies of all levels are obtained, the
wavelengths for all transitions are calculated consistently using the respective
upper ($E_u$) and lower ($E_l$) energy levels as $\nu_{ul} = (E_u-E_l)/h$.  This
is important in order to guarantee that we recover the correct local
thermodynamic equilibrium solution at high redshifts \citep[see discussion in
Sec.~3.2.1 of ][]{Jose2006}.

\subsubsection{Two-photon decay and non-dipole transitions}
Apart from the aforementioned transitions, we also include in our computations
the two photon decays of the $\HeIlevel{2}{1}{S}{0}$ and
$\HeIlevel{2}{3}{S}{1}$ levels. We adopt the values
$A_{\HeIlevel{2}{1}{S}{0}-\HeIlevel{1}{1}{S}{0}} = 51.3$~s$^{-1}$ and
$A_{\HeIlevel{2}{3}{S}{1}-\HeIlevel{1}{1}{S}{0}} = \pot{4.09}{-9}$~s$^{-1}$
\citep{DrakeVictor1969}.

Our final spectrum also contains the contribution of the
$\HeIlevel{2}{1}{S}{0}$ two-photon decay spectrum, which is computed like for
the hydrogen case \citep[see e.g. Eq.~3 in][]{Jose2006}. The fit to the
profile function \citep{Drake1986} for this transition is taken from
\cite{Switzer2007I}.

We also included some additional low probability non-dipole transitions
\citep{Lach2001}, but in agreement with previous studies found them to be
negligible.

\subsection{$\ion{He}{ii}$ model atom}
For singly ionized helium we use hydrogenic formulae \citep[see
e.g.][]{Jose2006} with re-scaled transition frequencies \citep[see
also][]{Switzer2007I}.  
The 2s two-photon decay profile is modelled using the one for hydrogen, adopting
the (re-scaled) value of $A^{\rm HeII}_{2{\rm s}\rightarrow 1{\rm s}} =
526.5$~s$^{-1}$.

%
\section{Inclusion of the hydrogen continuum opacity}
\label{sec:HIopacity}
In order to include the effect of absorption of photons close to the optically
thick resonant transitions of helium during
$\ion{He}{ii}\rightarrow\ion{He}{i}$ recombination due to the presence of
neutral hydrogen, one has to study in detail how the photons escape in
the helium lines. This problem has been solved by two of us using a diffusion
code, and the results will be presented in a separate paper (Chluba \& Sunyaev
2008, in preparation).

For the purposes of this paper, the important conclusion is that the results
obtained using this diffusion code are in rather good agreement with those
presented in \citet{Switzer2007I}, showing that for the interaction of photons
with the considered resonances, the hypothesis of {\it complete redistribution}
of the photons over the Voigt profile, $\phi(x)$ (see Appendix~\ref{app:Voigt}
for definitions), is {\it not correct}, and may lead to significant differences,
for the \HeILya~transition. However, that is not the case for the \HeIInt~line,
where the real dynamics is very close to the full redistribution case.
\subsection{The escape probability in the $\ion{He}{i}$ lines}
For the computations of the $\ion{He}{ii}\rightarrow\ion{He}{i}$ spectrum in
this paper, we make an ansatz about the shape of the escape probability in these
lines, which is described below. This hypothesis permit us to efficiently
compute the escape probability in our codes, without reducing the computational
time significantly. This ansatz has been tested against the full results (Chluba
\& Sunyaev 2008, in preparation), and is found to produce accurate results for
the spectrum.
To present it, we first describe two particular ``limiting'' cases for the
escape problem: the {\it complete redistribution} (or incoherent scattering),
and the {\it no redistribution} case (or coherent scattering in the lab
frame). The realistic case will be referred as {\it partial redistribution} in
the line (or coherent scattering in the rest frame of the atom).

For these two particular cases, we follow the procedure outlined in
\citet{Switzer2007I}. It is based on the assumption that within the considered
range of frequencies around a given resonance a solution of the photon field,
including resonant scattering and hydrogen continuum absorption, can be obtained
under {\it quasi-stationary} conditions.  Within $\pm 1\%-10\%$ of the line
center (or roughly $\pm 10^3-10^4$ Doppler width), this approximation should be
possible\footnote{In addition, the used approximation $\partial_\nu \nu
N_\nu\approx \nu_0 \partial_\nu N_\nu$, where $\nu_0$ is the transition
frequency of the considered resonance and $N_\nu=I_\nu/h\nu$, demands that the
obtained solution is only considered sufficiently close to the line center. For
this reason the term connected with emission of photon due to the recombination
of hydrogen (see definition of $I_{C}$ in \citet{Switzer2007I}) should be
neglected, since in this process practically all photons are emitted very close
to the ionization frequency $\nucH\ll\nu_0$ of hydrogen.}.

\subsubsection{ Complete redistribution (or incoherent scattering) }
One finds that in this case, the corresponding correction, $\Delta P_{\rm esc}$,
to the standard Sobolev escape probability, $P_{\rm S}=[1-e^{-\tau_{\rm
S}}]/\tau_{\rm S}$, is given by:
\begin{align}
\label{eq:DPesc}
\Delta P_{\rm esc}=\int_{-\infty}^{\infty} \phi(x) \id x \int_x^\infty
\!\!\!\tauS\,\phi(x') \, e^{-\tau_{\rm L}(x, x')} \Big[1-e^{-\tau_{\rm c}(x,
x')}\Big] \id x'.
\end{align}
Here $\tauL(x, x')=\tauS[\chi(x')-\chi(x)]$ is the optical depth with respect to
line scattering off the resonance, where $\tau_{\rm S}$ is the Sobolev optical
depth and $\chi(x)=\int_{-\infty}^x \phi(y) \id y$ is the normalized
($\chi(+\infty)=1$) integral over the Voigt-profile. Furthermore we introduced
the hydrogen continuum optical depth
\begin{subequations}
  \label{eq:tau_c_all}
  \begin{align}
    \label{eq:tau_c}
    \tau_{\rm c}(x, x') 
    &= \frac{c\, N^{\rm H}_{\rm 1s} }{H}\int_{\nu(x)}^{\nu(x')} 
    \sigma^{\rm H}_{\rm 1s}(\tilde{\nu}) \frac{\id \tilde{\nu}}{\tilde{\nu}}
    \\
    \label{eq:tau_c_appr}
    &\approx
    \frac{c\, N^{\rm H}_{\rm 1s} \sigma^{\rm H}_{\rm
        1s}(\nu)}{H \,\nu}\times\frac{\nu}{3}\left[1-\left(\frac{\nu}{\nu'}\right)^3\right], 
  \end{align}
\end{subequations}
where $\nu(x)=\nu^{\rm He}_0+x\,\Delta\nu_{\rm D}$, $N^{\rm H}_{\rm 1s}$ is the
number density of hydrogen atoms in the 1s-state, $\sigma^{\rm H}_{\rm 1s}(\nu)$
is the photoionization cross section of the hydrogen ground state, $H$ is the
Hubble expansion factor, and $\nu^{\rm He}_0$ is the transition frequency of the
considered helium resonance. The Doppler width, $\Delta\nu_{\rm D}$, of the line
due to the motion of helium atoms is defined in Appendix~\ref{app:Voigt}.

The computational details about the numerical integration of Eq.~\ref{eq:DPesc},
as well as the derivation of a one-dimensional integral approximation to the
full 2-dimensional integral are discussed in Appendix~\ref{app:pesc}.

\subsubsection{ No redistribution (or coherent scattering in the lab frame) }
This case corresponds to a situation in which every photon
coming through the line is emitted again with the same frequency. During
recombination this is a very good approximation in the very distant
damping wings of the resonance. In practise, this case can be treated using
the formalism described in \citet{Switzer2007I}. For every transition $u
\rightarrow l$, we need to define the following quantity
\begin{equation}
f_{u\rightarrow l} = \frac{R^{\rm out}_{u\rightarrow l}}{ A_{u\rightarrow l} +
  R^{\rm out}_{u\rightarrow l}}
\end{equation}
where $R^{\rm out}_{u\rightarrow l}$ is the sum of the rates of all the possible
ways of leaving the upper level but excluding the considered resonance,
i.e. $R^{\rm out}_{u\rightarrow l} = R_{u \rightarrow c} + \sum_{i, i\ne l} R_{u
\rightarrow i}$. In this equation, we have introduced the (bound-bound) rates,
which are computed as
\begin{equation}
R_{u \rightarrow i} = 
\begin{cases}
A_{u\rightarrow i} [1 + n_{\rm bb}(\nu_{ui})],     &\text{$E_i < E_u$} \\
A_{i\rightarrow u} (g_i/g_u) n_{\rm bb}(\nu_{iu}),  &\text{$E_i > E_u$} \\
\end{cases}
\end{equation}
and the photoionization rate, $R_{u \rightarrow c}$.
This quantity, $f_{u \rightarrow l}$, gives the fractional contribution to the
overall width of the upper level of all possible transitions leaving the upper
level except for the resonance. In other words, $f_{u\rightarrow l}$ represents
the branching fraction for absorption of a line photon to result in incoherent
scattering. During helium recombination $f_{u \rightarrow l}\sim 10^{-3}$ for
the \HeILya transition and close to unity for \HeIInt~line.

Once we have obtained this quantity for the considered transition, the
corresponding escape probability in the case of fully coherent scattering is
given by
\begin{equation}
P_{\rm esc} = \frac{f_{u\rightarrow l} P  }{1 - (1-f_{u\rightarrow l})P}
\label{eq:Pesc_noscatt}
\end{equation}
where
\begin{equation}
P = P_{\rm S}( f_{u\rightarrow l} \tau_{\rm S} ) + 
\Delta P_{\rm esc}( f_{u\rightarrow l} \tau_{\rm S}, \tau_{\rm c} )
\end{equation}
where $P_{\rm S}( f_{u\rightarrow l} \tau_{\rm S} )$ means that the Sobolev
escape probability is evaluated at $f_{u\rightarrow l} \tau_{\rm S}$ instead of
at $\tau_{\rm S}$; and obtaining $\Delta P_{\rm esc}(f_{u\rightarrow l}
\tau_{\rm S}, \tau_{\rm c})$ reduces to the use of equation~\ref{eq:DPesc}, but
evaluating it at $f_{u\rightarrow l} \tau_{\rm S}$ instead of at $\tau_{\rm S}$,
while $\tau_{\rm c}$ remains unchanged.

\begin{figure}
  \centering
  \resizebox{0.95\hsize}{!}{\includegraphics{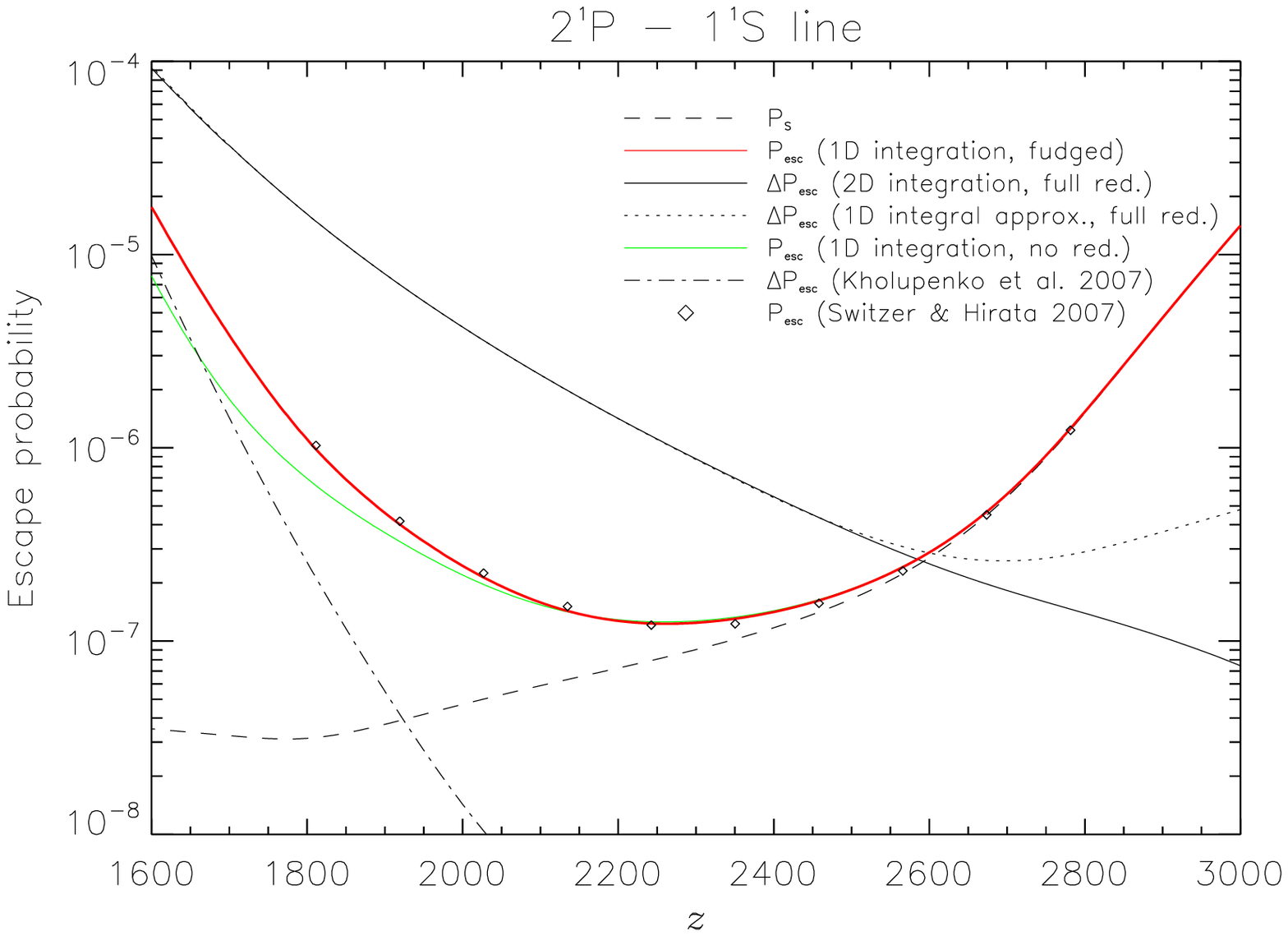}}
\\
  \resizebox{0.95\hsize}{!}{\includegraphics{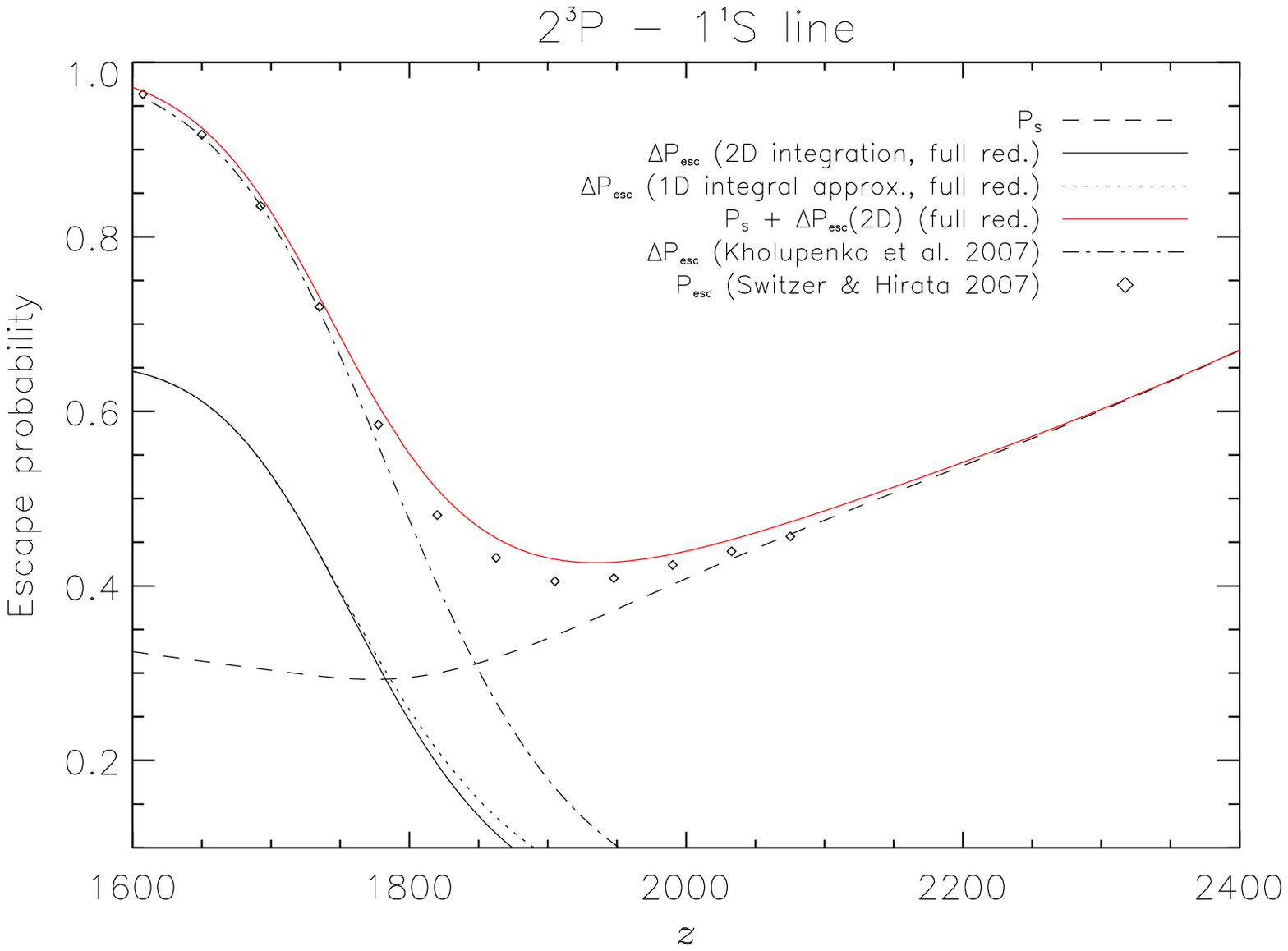}}
  \caption{ Contributions to the escape probability of the \HeILya transition
    (upper panel) and the \HeIInt-line (lower panel).
    $P_{\rm S}$ denotes the standard Sobolev escape probability. The correction
    to the escape probability due to the hydrogen continuum opacity is shown in
    several approaches: (a) full redistribution case: the full 2D-integral,
    $\Delta P_{\rm 2D}$, as given by Eq. \eqref{eq:DPesc}, and the 1D
    approximation, $\Delta P_{\rm 1D}$, obtained with
    Eq. \eqref{eq:DPesc_chi_1D}; (b) no redistribution case: the 1D
    approximation $P_{\rm 1D}$; and (c) partial redistribution case: the fudged
    escape probability based on the 1D integration of the no redistribution
    case.
    For comparison we show the simple analytic approximations of
    \citet{Kholupenko2007}, and the points extracted from Fig.~11 of
    \citet{Switzer2007I} for the case labelled as coherent (which is the
    equivalent to our partial redistribution case) for the upper panel, and the
    points extracted from Fig.~4 of \citet{Switzer2007III}.}
\label{fig:Pesc}%
\end{figure}

\subsubsection{ Partial redistribution case. Our ansatz}
The detailed treatment of the problem with {\it partial redistribution} is
computationally demanding. Our results are based on a diffusion code (Chluba \&
Sunyaev 2008, in preparation), which requires $\sim 1$~day on a single $3\,$GHz
processor to treat one cosmology. The other approach of this problem, based on a
Monte Carlo method \citep{Switzer2007I} is equally demanding.

For the computations of this paper, we propose and test an ansatz which permit
us to compute efficiently the escape probability.
Our basic assumption is that {\it the ratio of the escape probability in the
complete problem (partial redistribution case) to the escape probability in the
problem with no redistribution is a constant number for a given redshift, or
equivalently, it has a very small dependence on the recombination history}.  In
that case, we can use this function (the ratio of those two cases) to {\it
fudge} the real escape probability in our code in a very fast way.
The important thing is that we only need to compute a single solution of the
complete problem in order to tabulate the fudge function.  Moreover, the
reference case (the no redistribution case) is fully analytic, and using our 1D
integral approximation described in Appendix~\ref{app:pesc}, it is obtained very
fast. Summarising, this scheme permit us to compute the escape probability with
high accuracy in our codes, without the need of interpolating using pre-computed
tables.

In practise, we use the solution for the recombination history which was
obtained within the no redistribution approximation, and we compute the
corresponding escape probability for a given cosmology. If we make a further
iteration, by recomputing the new recombination history using the new escape
probability, we find that the result practically does not change.
For the \HeIInt-line escape probability is always close to the full
redistribution case, so for this line we directly consider this approximation
for the computations.
%

\subsection{Results for $\Delta P_{\rm esc}$}
\label{sec:DPesc_res}
In this paper, we only consider the corrections to the escape probability for
the \HeILya transition and \HeIInt-line.
In principle, all the other $\HeIlevel{n}{1}{P}{1}-\HeIlevel{1}{1}{S}{0}$ and
spin-forbidden transitions are also affected, by the presence of neutral
hydrogen, but the effect is smaller, and we omit these additional corrections
for the moment.

In Fig.~\ref{fig:Pesc} we show different contributions to the escape
probability of the two considered transitions, computed within
different approximations discussed in the last subsection. 
In our computations, for the \HeILya transition the effect of hydrogen is
starting to become important below $z\sim 2400-2500$, whereas in the case of
the \HeIInt-line the escape probability is strongly modified only at
$z\lesssim 1800-1900$.
One can also clearly see, as illustrated for the complete redistribution
approach, that in both cases the 1D-approximation works extremely well at nearly
all relevant redshifts.
In particular, the differences are small where the deviations between the inner
integrand in Eq.~\ref{eq:DPesc}, and its analytic approximation deduced
from Eq.~\eqref{eq:DPesc_chi_1D} are small (see Fig.~\ref{fig:examples.F}, and
the discussion in Appendix~\ref{app:pesc}).

For the \HeILya transition, the departure of the escape probability with respect
to the full redistribution case is very important, being at least one order of
magnitude different at redshifts below $z\approx 2200$. Moreover, the full
redistribution case becomes important at earlier redshifts ($z\sim 2600-2700$),
thus producing a recombination dynamics which would be much closer to the Saha
solution. In other words, the assumption of complete redistribution
significantly overestimates the escape rate of photons from the \HeILya
transition, and thus would artificially accelerate
$\ion{He}{ii}\rightarrow\ion{He}{i}$ recombination.
The final (fudged) solution is in reality much more close to the ``no
redistribution'' case, although the differences with respect to this later case
are still significant (roughly a factor of 2 at $z\sim 1730$).
Comparing our results with other recent computations, we find that our final
(fudged) solution is very close to the \citet{Switzer2007I} computation, which
was based on a Monte Carlo analysis of the escape problem.  There are still
small differences around redshifts $z\sim 2200-2400$, which could be probably
due to the fact that we do not include the modified escape for higher levels.
However the formula given in \citet{Kholupenko2007} only works at very low
redshifts.

For the \HeIInt-transition, the situation is different. The computations, based
on the diffusion code, show that for this line one can approximate the escape
probability using complete redistribution at the level of $\sim 10\%$.
Therefore, for the computations in this paper, we adopt this approximation for
this particular transition. 
The lower panel of Fig.~\ref{fig:Pesc} also shows the comparison between our
escape probability and those obtained in other recent publications. The
agreement with the \citet{Switzer2007I} result is again very remarkable, except
for the redshift region around $z\sim 1900$. However, we have checked that this
difference is mainly due to the assumption of using the full redistribution
solution for this line, and that this difference implies only small changes in
the final recombination spectrum.
For this transition the actual corrections due to electron scattering, which we
neglected so far, are larger.
Finally we note that, although in this case there is an apparent agreement at
low redshifts with the \citet{Kholupenko2007} result, their computation
corresponds to the quantity $\Delta P_{\rm esc}$. Thus, when adding the
contribution of the Sobolev escape, they have a value of the probability which
exceeds unity.

\subsection{Inclusion into the multi-level code}
\label{sec:inclusion2code}
In order to account for the effect of the hydrogen continuum opacity during
$\ion{He}{ii}\rightarrow\ion{He}{i}$ recombination into our multi-level code
several changes are necessary. The first and most obvious modification is the
replacement of the Sobolev escape probability $P_{\rm S}\rightarrow P_{\rm
S}+\Delta P_{\rm esc}$ for the \HeILya and \HeIInt-transition.
Due to the above replacement {\it more} electrons are reaching the ground state
of neutral helium, but {\it no additional} helium photons are released.
Therefore in the computation of the helium spectrum the increase in the photons
escape rate by $\Delta P_{\rm esc}$ should {\it not} be included.

Given the usual {\it net} radiative transition rate $P_{\rm S}\times\Delta
R_{i\rm 1^1S}$ from level $i$ to the helium ground state, the increase in the
net transition rate due to the presence of neutral hydrogen atoms is given by
$\Delta R^{\rm abs}_{i\rm 1^1S}=\Delta P_{\rm esc}\times \Delta R_{i\rm 1^1S}$.
Since the corresponding photons associated with this transition are ionizing
hydrogen atoms one has to add the rate $\Delta R^{\rm abs}_{i\rm 1^1S}$ to the
electron equation and subtract it from the hydrogen 1s-equation.
Although it is clear that, given this small addition of electrons to the
continuum, the hydrogen ground state population will re-adjust within a very
short time, it is still possible that the corresponding electrons will reach the
ground state via various decay channels, including a cascade from highly excited
levels, which may even end in the 2s level, yielding two photons in the
two-photon decay transition.
Instead of assuming that {\it all} electrons connected with the increase of the
net transition rate, $\Delta R^{\rm abs}_{i\rm 1^1S}$, are leading to the
emission of a hydrogen Lyman-$\alpha$ photon {\it only} \citep[as done in
][]{Kholupenko2007}, this approach is more consistent.
We shall see below that a part of the additional electrons indeed take more
indirect routes to the hydrogen 1s-level.

With these additions to our multi-level code it is possible to obtain both the
ionization history and the helium and hydrogen recombination spectrum including
the effect of the hydrogen continuum opacity as outlined in this Section.

\begin{figure}
  \centering \resizebox{\hsize}{!}{\includegraphics{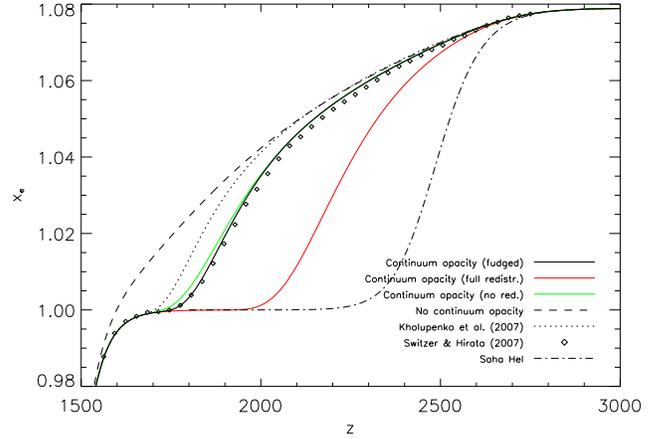}}
  \caption{The ionization history during the epoch of
  $\ion{He}{ii}\rightarrow\ion{He}{i}$ recombination for different
  approaches.}
\label{fig:xe}
\end{figure}

%
\section{The helium recombination history}
\label{sec:Xe}
The main goal of this paper is to compute the spectral distortions resulting
from the bound-bound transitions of helium. However, since we are discussing
several approximations to include the hydrogen absorption during the epoch of
$\ion{He}{ii}\rightarrow\ion{He}{i}$ recombination, we here shortly discuss the
corresponding differences in the ionization fraction.

Figure~\ref{fig:xe} shows our results for the redshift-dependence of the free
electron fraction $x_{\rm e} = n_{\rm e} / n_{\rm H}$ during
$\ion{He}{ii}\rightarrow\ion{He}{i}$ recombination, using the three
approximations for the escape probability in the $\ion{He}{i}$ lines, as
discussed above.
Qualitatively, all three results (i.e. full redistribution, no
redistribution and partial redistribution) agree with those found in some
earlier studies \citep{Kholupenko2007,Switzer2007I}, showing that the inclusion
of the hydrogen continuum opacity in the computation significantly speeds up
recombination, making it closer to the Saha solution.
Our (fudged) solution for the case of partial redistribution of photons in the
resonance is in good agreement with the \citet{Switzer2007I}, except for the
small difference around $z\sim 2200$. As pointed out in the last section, these
are likely due to the fact that we did not include the continuum opacity
correction for higher transitions.

It is important to note that the incorrect hypothesis of full redistribution in
the \HeILya-resonance has a strong impact on the recombination history. In that
case, the effect of continuum opacity on the escape probability becomes of
importance at earlier times, shifting the redshift at which $x_{\rm e}$ starts
to depart from the solution without continuum opacity considerably. In addition,
the period during which $x_{\rm e}$ is very close to unity, i.e. just before
hydrogen recombination starts, becomes considerably longer.
Therefore, it is very important for a detailed analysis of the recombination
history to treat properly the escape probability in this line.

So far we did not consider the effect of feedback in our computations and as
shown in \citet{Switzer2007I} one does expect some additional delay of
$\ion{He}{ii}\rightarrow\ion{He}{i}$ recombination around $z\sim 2400$.
However, looking at Fig. 12 in \citet{Switzer2007I}, this process is not
expected to alter the results by more than $10\%-20\%$.

Finally, we also mention that for computations of the electron fraction during
the epoch of $\ion{He}{ii}\rightarrow\ion{He}{i}$ recombination, it is {\it not}
necessary to include a very large number of shells.
Unlike in the case of hydrogen recombination, the exponential tail of
$\ion{He}{ii}\rightarrow\ion{He}{i}$ recombination, which potentially is the
most sensitive to the completeness of the atomic model, is entirely buried by
the large number of free electrons from hydrogen.
In addition, practically no ionized helium atoms remain after recombination,
although in the case of hydrogen a small residual fraction remains.
This is because there are significantly more electrons per helium atom than
for hydrogen, such that freeze-out for helium occurs at an exponentially lower
level.

Our results suggest that for $\ion{He}{ii}\rightarrow\ion{He}{i}$ recombination
the inclusion of 5 shells is already enough to capture the evolution of $x_{\rm
e}$ during this epoch with precision better than 0.1~\%.
This precision is sufficient if one is interested in cosmological parameter
estimation from the angular power spectra ($C_\ell$'s) of the CMB.
However, still rather significant modifications of the recombination history can
be expected in particular from the feedback of $\ion{He}{ii}$-photons and
probably other physical processes that were omitted here \citep[e.g.
see][]{Switzer2007I}.
%

%
\begin{figure*}
\centering 
\includegraphics[height=1.7\columnwidth,width=1.2\columnwidth,angle=90]{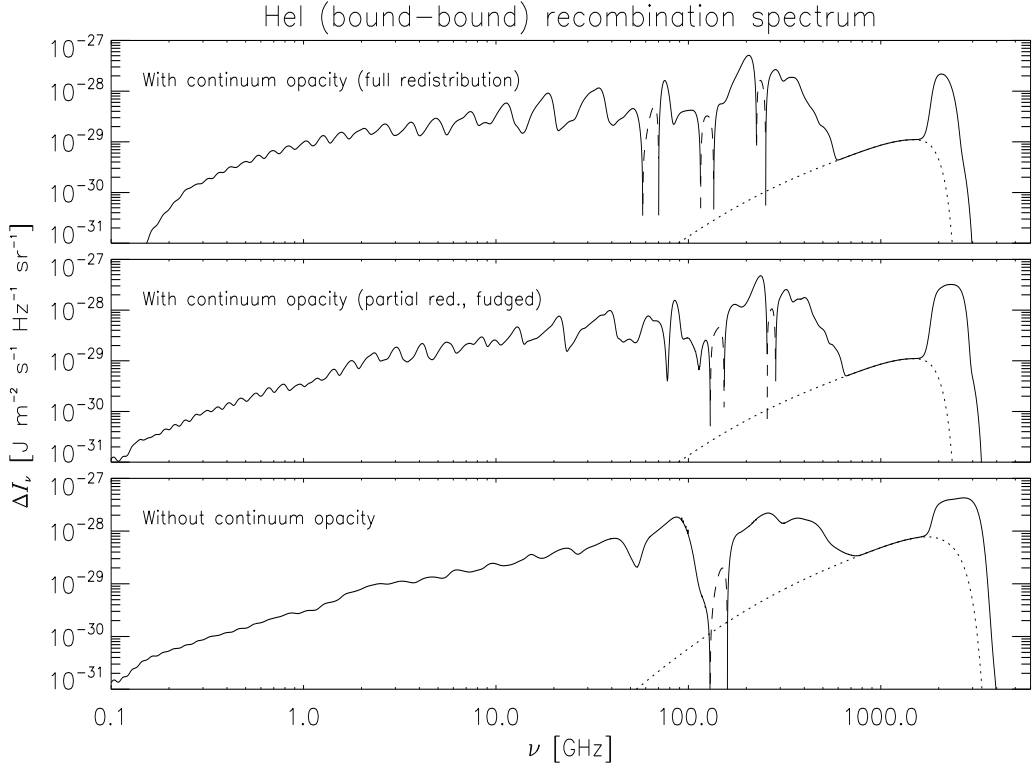}
\caption{ A comparison of the $\ion{He}{i}$ recombination spectrum for
  $\nmax=30$ with (top and middle panels) and without (bottom panel) the
  inclusion of the effect of hydrogen continuum opacity. Upper panel corresponds
  to the case of full redistribution of photons in the \HeILya -resonance, while
  the middle panel corresponds to our final (fudged) computation for the partial
  redistribution case (see text for details). Ordered in this way, from top to
  bottom we have progressively a slower recombination, so all the features
  become broader. In all three cases, a solid line indicate positive values,
  while a dot-dashed line indicates negative ones. Two-photon decay continuum
  2$^1$S$_0$ - 1$^1$S$_0$ is also included as dotted line in both panels. }
  \label{fig:compare}%
\end{figure*}

\section{Bound-Bound helium recombination spectra}
\label{sec:He.spectra}
In Fig.~\ref{fig:final} we present the main result of this paper, namely the
complete bound-bound helium recombination spectrum, arising both during the
epoch of $\ion{He}{iii} \rightarrow \ion{He}{ii}$ ($5000\lesssim z\lesssim
7000$), and $\ion{He}{ii} \rightarrow \ion{He}{i}$ recombination ($1600\lesssim
z\lesssim 3000$).
For comparison, we also included the results obtained in our previous
computations for the \ion{H}{i} bound-bound recombination \citep{Chluba2007},
and added the additional line appearing as a consequence of the re-processing
of \ion{He}{i} photons in the continuum of hydrogen, as described below (see
Sect.~\ref{sec:comp_H_abs.H.spec}).
Also the 2s two-photon decay continua for all cases are shown.
There are two important issues to be mentioned here:
\begin{enumerate}
\item[(i)] First, the helium spectral features (both for \ion{He}{ii} and
\ion{He}{i}) are significantly narrower than those of the hydrogen recombination
spectrum.
This is due to the fact that for \ion{He}{i}, recombination occurs significantly
faster due to the inclusion of the hydrogen continuum opacity, and in the case
of \ion{He}{ii}, because its recombination occurs much more close to Saha
conditions in the first place.
Even the inclusion of Doppler broadening due to electron scattering, as
described in Sect.~\ref{sec:spec.e-scatt}, is unable to change this aspect.
As a consequence, both recombination spectra contain clear features in the low
frequency domain ($\nu \sim 1$~GHz), where the \ion{H}{i} spectrum is
practically featureless. This increase in the amplitude of variability of the
recombinational radiation at low frequencies might help to detect these features
in the future.
\item[(ii)] Secondly, the \ion{He}{i} recombination spectrum displays {\it two
negative features}, at positions $\nu \approx 145$ and $270$~GHz. This is
qualitatively different from the case of the hydrogen and \ion{He}{ii} spectra,
where the net bound-bound spectra appear in emission. As we will discuss below,
the reason for these features is directly connected with the dynamics of
recombination. They are associated with transitions in which the lower state is
effectively ``blocked'' for all downward transitions, such that faster channels
to the $\HeIlevel{1}{1}{S}{0}$ level are provided through energetically higher
levels.
\end{enumerate}
We now discuss in detail some particular aspects of the recombination spectra.

%
\subsection{Importance of the hydrogen continuum opacity for the bound-bound 
\ion{He}{i} spectrum}
\label{sec:comp_H_abs.He.spec}
In Fig.~\ref{fig:compare} we show the comparison between the \ion{He}{i}
spectrum in three cases, namely the case {\it with} hydrogen continuum opacity
assuming full redistribution of photons in the resonance; the case {\it with}
hydrogen continuum opacity assuming partial redistribution; and the case {\it
without} the inclusion of the hydrogen continuum opacity in the computation.

The width of the lines, which is directly connected with the duration of the
recombination process, is significantly smaller when including the effect of
hydrogen continuum opacity. In addition, the peaks of the lines are shifted
towards lower frequencies (i.e. higher redshifts).  As a consequence, the
spectrum has a richer structure as compared to the case of hydrogen, since the
overlap of lines is smaller.

The full redistribution and partial redistribution spectra are very similar in
the low frequency ($\nu \la 30$~GHz) region.
However, at higher frequencies, several differences appear. In particular, for
the full redistribution computation there are {\it three} negative features
instead of two.
The spectrum for the case without continuum opacity is much smoother than the
previous two, and presents only {\it one} negative feature.
In addition, in this frequency regime (and specially for the strong feature at
$\nu \ga 2000$~GHz) is seen that, due to the different speeds of the
recombination process, the lines appear displaced towards lower frequencies
(higher redshifts) as we move from the lower to the upper panel.

\begin{figure}
  \centering
  \resizebox{\hsize}{!}{\includegraphics{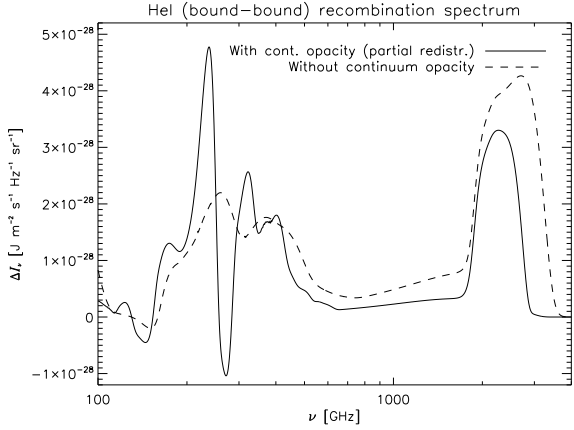}}
  \caption{ A comparison of the \ion{He}{i} recombination spectrum for
    $\nmax~=~30$ close to the \HeILya line, with (solid curve) and without
    (dashed curve) the effect of hydrogen continuum opacity. The case with
    continuum opacity corresponds to the partial redistribution (fudged)
    computation. The two-photon decay continuum 2$^1$S$_0$ - 1$^1$S$_0$ is also
    included in the spectra. }
  \label{fig:compare_zoom}%
\end{figure}

\begin{figure}
  \centering
  \resizebox{\hsize}{!}{\includegraphics{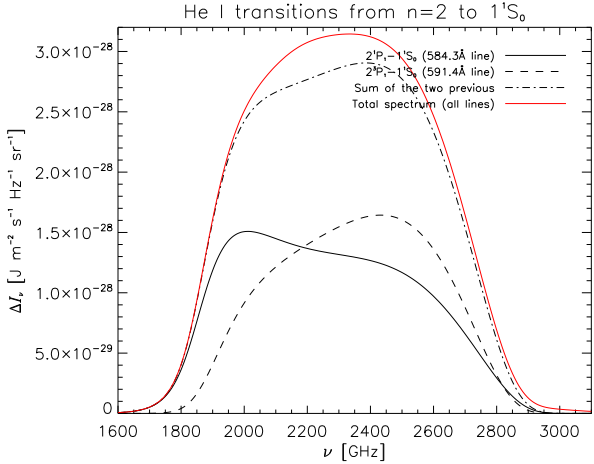}}
  \caption{Transitions in \ion{He}{i} atom from the $n=2$ shell to the ground
  state, in particular the \HeILya and \HeIInt-lines.
This figure was obtained using our results for the $\nmax=20$ computation,
including the effect of hydrogen continuum opacity in the partial redistribution
case (fudged solution).}
\label{fig:lya}%
\end{figure}

For the high frequency region ($\nu > 100$~GHz), we present a more detailed
direct comparison in Fig.~\ref{fig:compare_zoom} between the cases of partial
redistribution and the one without continuum opacity. In this figure, a linear
scale in the vertical axis is used in order to emphasise the existence of the
negative features.
One can see that the relative contribution of the different lines is strongly
altered. In general, all emission appearing above $\nu\sim500\,$GHz
is suppressed, while at lower frequencies, some lines are enhanced.
We can understand these changes as follows: 
the contributions appearing at $\nu \gtrsim 500$~GHz correspond to the
$\HeIlevel{n}{1}{P}{1}-\HeIlevel{1}{1}{S}{0}$-series of neutral helium, the
spin-forbidden transitions directly connecting to the ground state
($\HeIlevel{n}{3}{P}{1}-\HeIlevel{1}{1}{S}{0}$), and the two-photon decay of
the ${\HeIlevel{2}{1}{S}{0}}$ singlet-state.
The first two series contribute most to the strong feature at $\nu \gtrsim
2000$ (see Fig.~\ref{fig:lya} for some more detail), while the broad
two-photon continuum dominates the spectrum in the vicinity of
$\nu\sim1000$~GHz.
When the hydrogen continuum opacity is included, in our current implementation
of the problem, only the \HeILya and \HeIInt-transition are {\it directly}
affected, i.e. via the inclusion of $\Delta P_{\rm esc}$, whereas all the
other lines are modified only {\it indirectly} because of to the change in the
{\it recombination dynamics} and the relative importance of different escape
channels.

Figure~8 of \citet{Wong2007} shows that without the hydrogen
continuum opacity the \HeILya channel defines the rate of recombination at
$z\gtrsim 2400$, while at $z\lesssim 2400$ the $\HeIlevel{2}{3}{P}{1} -
\HeIlevel{1}{1}{S}{0}$ spin-forbidden and, to a smaller extent, the
$\HeIlevel{2}{1}{S}{0}$ two-photon decay channel dominate.
They computed that of all electrons that reach the ground state of helium
$39.9\%$ go through the \HeILya transition, $42.8\%$ pass through the
$\HeIlevel{2}{3}{P}{1} - \HeIlevel{1}{1}{S}{0}$ spin-forbidden transition and
only $17.3\%$ take the route via the $\HeIlevel{2}{1}{S}{0}$ two-photon decay
channel.

In our computations including the hydrogen continuum opacity we have to keep in
mind that there is a fraction of electrons that reach the helium ground due to
continuum absorption by hydrogen, which then lead to the emission of additional
photons in the \ion{H}{i} recombination spectrum.
Direct integration of the total number of photons in the neutral helium spectrum
around $\nu\sim 2000$~GHz yields $N_\gamma = 4\pi/c \int d\nu \Delta I_\nu /
(h\nu) \sim 0.46 N_{\rm He}$, while for the $\HeIlevel{2}{1}{S}{0}$ two-photon
decay spectrum we have $\sim 0.16 N_{\rm He}$ photons.
Moreover, one can compute the number of photons in the newly generated hydrogen
Ly$\alpha$ line (see Figure~\ref{fig:lya_hyd} below), obtaining $N_\gamma ({\rm
Ly}\alpha) \sim 0.44 N_{\rm He}$.
These numbers show that $\sim 90\%$ of all electrons that reach the ground state
of helium pass through the \HeILya and \HeIInt-transition. The
$\HeIlevel{2}{1}{S}{0}$ two-photon channel only allows $\sim 8\%$ of all helium
atoms to recombined, and $\sim 2\%$ go through the other $\HeIlevel{n}{1}{P}{1}
- \HeIlevel{1}{1}{S}{0}$ and $\HeIlevel{2}{3}{P}{1} - \HeIlevel{1}{1}{S}{0}$
spin-forbidden transitions.
We note that the modification of the dynamics of \ion{He}{i} recombination is
influencing the relative amplitude of other lines, such as the $3^1{\rm D} -
2^1{\rm P}$ ($6680~\AA$) transition, which is strongly amplified, or the
$3^3{\rm D} - 2^3{\rm P}$ ($5877~\AA$) transition, which now appears in
absorption. We will discuss these transitions in detail below.

To end this subsection, we remind again that those features in the vicinity of
$\nu \sim 2000$~GHz arising from the \HeILya and \HeIInt-transitions will not be
observed in the real spectrum, because of feedback processes connected with HI
continuum absorption at lower redshifts will take away these photons and will
produce additional distortions in the HI spectrum.

\subsection{Importance of the hydrogen continuum opacity for the 
bound-bound \ion{H}{i} spectrum}
\label{sec:comp_H_abs.H.spec}
\begin{figure}
  \centering \resizebox{\hsize}{!}{\includegraphics{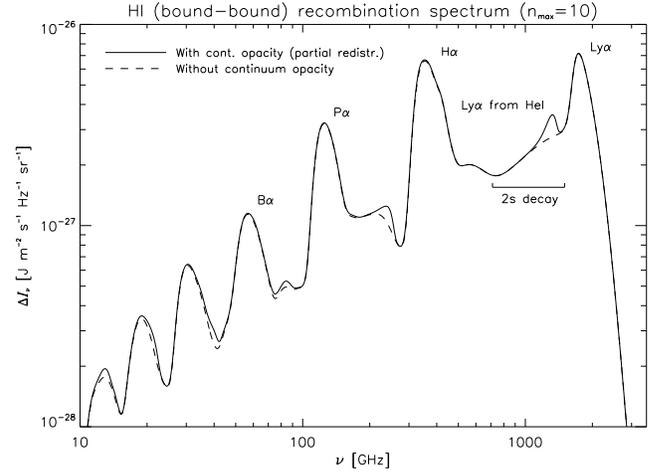}}
        \caption{ A comparison of the \ion{H}{i} recombination spectrum for
        $\nmax=10$ at high frequencies, with (solid curve) and without (dashed
        curve) the hydrogen continuum opacity was included in the treatment of
        the \ion{He}{i} atom. The 2s two-photon decay continuum is also shown in
        both cases. }
\label{fig:compare_hyd}%
\end{figure}

\begin{figure}
  \centering
  \resizebox{\hsize}{!}{\includegraphics{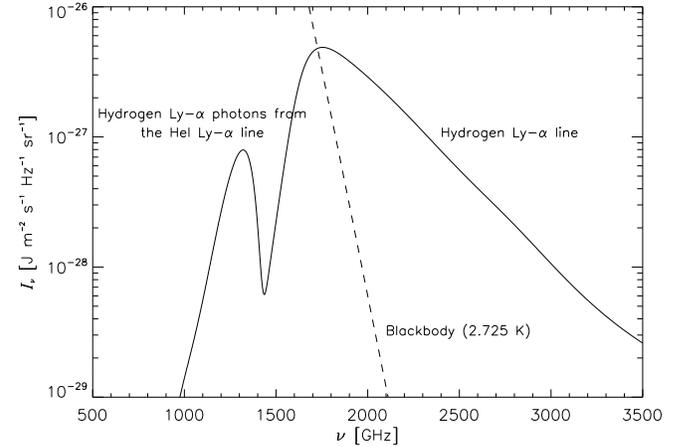}}
  \caption{ Hydrogen Ly-$\alpha$ recombinational line. When including the effect
  of continuum opacity and solving simultaneously the evolution of hydrogen and
  helium recombination, most of the electrons which are taken from the helium by
  neutral hydrogen atoms re-appear as a hydrogen Ly-$\alpha$ line at $\nu\approx
  1320$~GHz ($z\approx 1870$). }
  \label{fig:lya_hyd}%
\end{figure}

In Fig.~\ref{fig:compare_hyd} we show how the hydrogen recombination spectrum is
modified because of the additional free electrons appearing due to the
absorption of \ion{He}{i}-photons in the hydrogen continuum absorption.
Most of them recombine after a very short time through the main channel of
hydrogen recombination at high redshifts, which is the Ly-$\alpha$ transition
\citep[e.g. see Fig.~10 in ][]{Jose2006}, producing a ``new'' hydrogen
Ly-$\alpha$ feature at $z\approx 1870$. In Fig.~\ref{fig:lya_hyd} we show the
shape of that line separately.
However, as explained in Sect. \ref{sec:inclusion2code}, the small addition of
electrons to the continuum also produces changes in the rest of the hydrogen
spectrum, as shown in Figure~\ref{fig:compare_hyd}.  In some cases (see e.g. the
high-frequency wing of the Paschen series) the changes are important at the
level of 10 percent.
This feature is due to the new H$\alpha$ feature.
On average, the new bound-bound \ion{H}{i} spectrum is slightly higher in
amplitude, as a consequence of the additional photons appearing in this
process. Because the re-processing of \ion{He}{i} photons occurs at high
redshifts (above $z=1800$), the two-photon continuum line is practically
unchanged.

%
\begin{table}
\caption{Positions and amplitudes of all the negative lines in the \ion{He}{i}
recombination spectrum for $\nmax=20$ with peak intensities smaller than
$\pot{-1}{-29}$~${\rm J\,m^{-2}\,s^{-1}\,Hz^{-1}\,sr^{-1}}$. We show the
corresponding terms for the lower and upper states, the peak intensity at the
minimum, the central frequency ($\nu_0$) as observed today, the redshift
($z_{\rm min}$) at the minimum, and the wavelength for that transition in the
rest frame ($\lambda_{\rm rest}$).}
\label{table:1}
\centering     
\begin{tabular}{c c c c c c} 
\hline\hline            
Lower & Upper & $\Delta I_\nu$ at minimum  & $\nu_0$ & $z_{\rm min}$ & $\lambda_{\rm rest} $\\
 & & [${\rm J\,m^{-2}\,s^{-1}\,Hz^{-1}\,sr^{-1}}$] & [GHz] & & [\AA] \\
\hline        
$\rm 2 ^ 1S_ 0$ & $\rm 2 ^ 1P_ 1$ & $\pot{-1.1}{-28}$ &  78 & 1855 & 20590 \\
$\rm 3 ^ 3D_ 2$ & $\rm 4 ^ 1F_ 3$ & $\pot{-1.1}{-28}$ &  86 & 1875 & 18690 \\
$\rm 3 ^ 3D_ 3$ & $\rm 4 ^ 1F_ 3$ & $\pot{-1.5}{-29}$ &  86 & 1875 & 18690 \\
$\rm 3 ^ 3D_ 2$ & $\rm 5 ^ 1F_ 3$ & $\pot{-1.2}{-29}$ & 125 & 1875 & 12790 \\
$\rm 4 ^ 3D_ 2$ & $\rm 5 ^ 1F_ 3$ & $\pot{-1.2}{-29}$ &  40 & 1875 & 40380 \\
$\rm 2 ^ 3S_ 1$ & $\rm 2 ^ 3P_ 1$ & $\pot{-2.1}{-28}$ & 145 & 1905 & 10830 \\
$\rm 2 ^ 3P_ 1$ & $\rm 3 ^ 3D_ 2$ & $\pot{-1.6}{-28}$ & 273 & 1870 &  5877 \\
$\rm 2 ^ 3P_ 2$ & $\rm 3 ^ 3D_ 2$ & $\pot{-5.9}{-29}$ & 272 & 1875 &  5877 \\
$\rm 3 ^ 3D_ 2$ & $\rm 4 ^ 3F_ 3$ & $\pot{-1.0}{-28}$ &  86 & 1870 & 18690 \\
$\rm 3 ^ 3D_ 3$ & $\rm 4 ^ 3F_ 3$ & $\pot{-1.4}{-29}$ &  86 & 1875 & 18690 \\
$\rm 4 ^ 3D_ 2$ & $\rm 5 ^ 3F_ 3$ & $\pot{-1.1}{-29}$ &  40 & 1875 & 40380 \\
\hline\hline
\end{tabular}
\end{table}

\begin{table}
\caption{Positions and amplitudes of all the positive lines in the \ion{He}{i}
recombination spectrum for $\nmax=20$ with peak intensities greater than
$\pot{2}{-29}$~${\rm J\,m^{-2}\,s^{-1}\,Hz^{-1}\,sr^{-1}}$. We show the
corresponding terms for the lower and upper states, the peak intensity at the
maximum, the central frequency $\nu_0$ as observed today, and the wavelength
for that transition in the rest frame ($\lambda_{\rm rest}$).}
\label{table:2}
\centering     
\begin{tabular}{c c c c c c} 
\hline\hline            
Lower & Upper & $\Delta I_\nu$ at maximum  & $\nu_0$ & $z_{\rm max}$ & $\lambda_{\rm
  rest} $ \\
 & & [${\rm J\,m^{-2}\,s^{-1}\,Hz^{-1}\,sr^{-1}}$] & [GHz] &  & [\AA]  \\
\hline        
$\rm 1 ^1 S_ 0 $  &  $\rm 2 ^1 P_ 1 $ & $\pot{1.5}{-28}$ & 2011  & 2550 &   584.3 \\
$\rm 1 ^1 S_ 0 $  &  $\rm 2 ^3 P_ 1 $ & $\pot{1.6}{-28}$ & 2430  & 2085 &   591.4 \\
$\rm 2 ^1 S_ 0 $  &  $\rm 3 ^1 P_ 1 $ & $\pot{7.5}{-29}$ &  317  & 1885 &  5017 \\
$\rm 2 ^1 S_ 0 $  &  $\rm 4 ^1 P_ 1 $ & $\pot{2.2}{-29}$ &  401  & 1885 &  3966 \\
$\rm 2 ^1 P_ 1 $  &  $\rm 3 ^1 S_ 0 $ & $\pot{2.4}{-29}$ &  218  & 1885 &  7283 \\
$\rm 2 ^1 P_ 1 $  &  $\rm 3 ^1 D_ 2 $ & $\pot{4.6}{-28}$ &  239  & 1880 &  6680 \\
$\rm 2 ^1 P_ 1 $  &  $\rm 4 ^1 D_ 2 $ & $\pot{8.0}{-29}$ &  323  & 1885 &  4923 \\
$\rm 2 ^1 P_ 1 $  &  $\rm 5 ^1 D_ 2 $ & $\pot{3.0}{-29}$ &  362  & 1885 &  4389 \\
$\rm 3 ^1 D_ 2 $  &  $\rm 4 ^1 F_ 3 $ & $\pot{1.7}{-28}$ &   85  & 1875 & 18700 \\
$\rm 3 ^1 D_ 2 $  &  $\rm 5 ^1 F_ 3 $ & $\pot{3.1}{-29}$ &  125  & 1880 & 12790 \\
$\rm 3 ^1 D_ 2 $  &  $\rm 4 ^3 F_ 3 $ & $\pot{1.5}{-28}$ &   85  & 1875 & 18700 \\
$\rm 3 ^1 D_ 2 $  &  $\rm 5 ^3 F_ 3 $ & $\pot{2.1}{-29}$ &  125  & 1880 & 12790 \\
$\rm 4 ^1 D_ 2 $  &  $\rm 5 ^1 F_ 3 $ & $\pot{2.4}{-29}$ &   39  & 1880 & 40410 \\
$\rm 4 ^1 F_ 3 $  &  $\rm 5 ^1 G_ 4 $ & $\pot{2.3}{-29}$ &   39  & 1885 & 40490 \\
$\rm 2 ^3 S_ 1 $  &  $\rm 2 ^3 P_ 0 $ & $\pot{3.2}{-29}$ &  146  & 1890 & 10830 \\
$\rm 2 ^3 S_ 1 $  &  $\rm 2 ^3 P_ 2 $ & $\pot{8.4}{-29}$ &  142  & 1950 & 10830 \\
$\rm 2 ^3 S_ 1 $  &  $\rm 3 ^3 P_ 2 $ & $\pot{2.5}{-29}$ &  408  & 1890 &  3890 \\
\hline\hline
\end{tabular}
\end{table}

%
\subsection{Negative features in the \ion{He}{i} spectrum. }
One of the most interesting results of our computations is the existence of
two {\it negative} features in the \ion{He}{i} recombination spectrum.
In order to identify the transitions which contribute most to those features,
in Table~\ref{table:1} we provide a list with the position and amplitudes of
all the individual lines which are found to be negative at an amplitude
smaller than $\pot{-1}{-29}$~${\rm J\,m^{-2}\,s^{-1}\,Hz^{-1}\,sr^{-1}}$ from our
$\nmax=20$ computation.
To help in the discussion, we also provide in Table~\ref{table:2} a list with
the position and amplitudes of all the {\it positive} individual lines which are
found to have an amplitude larger than $\pot{1}{-29}$~${\rm
J\,m^{-2}\,s^{-1}\,Hz^{-1}\,sr^{-1}}$ from the same computation.

\begin{figure}
  \centering
  \resizebox{\hsize}{!}{\includegraphics{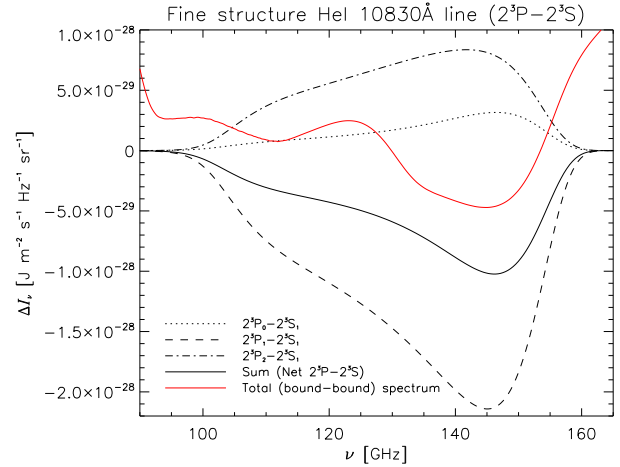}}
  \caption{ First negative feature in the \ion{He}{i} spectrum. The largest
  negative contribution is coming from the 10830~$\AA$ fine-structure line.
  Note that the upper level has fine structure, so we present all three possible
  values of $J$. This figure was obtained using our results for the $\nmax=20$
  computation, including the effect of hydrogen continuum opacity. }
  \label{fig:line1}%
\end{figure}

\begin{figure}
  \centering
  \resizebox{\hsize}{!}{\includegraphics{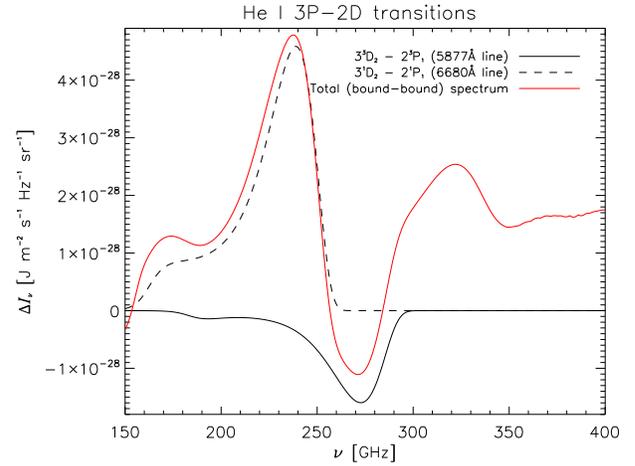}}
  \caption{ Second negative feature in the HeI spectrum. The largest negative
  contribution is produced by the 3D-2P (triplet-triplet) transition. This
  figure was obtained using our results for the $\nmax=20$ computation,
  including the effect of hydrogen continuum opacity.}
  \label{fig:line2}%
\end{figure}

\begin{figure}
  \centering
  \resizebox{\hsize}{!}{\includegraphics{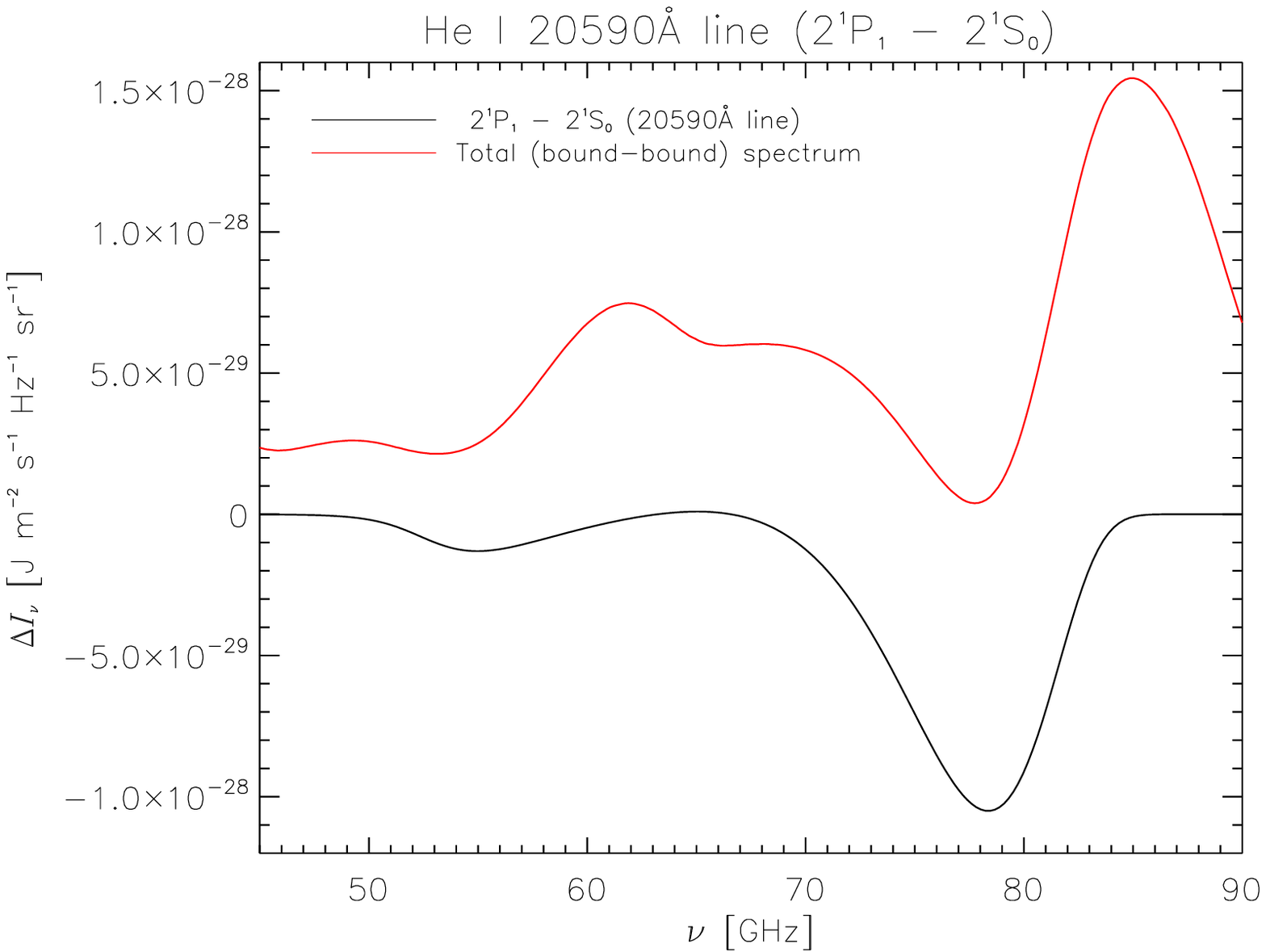}}
  \caption{ The \ion{He}{i} recombination spectrum in the vicinity of the
  $20590~\AA$ line. This figure was obtained using our results for the
  $\nmax=20$ computation, including the effect of hydrogen continuum opacity. }
  \label{fig:line3}%
\end{figure}

We now discuss in detail each one of these two negative features.
Figures~\ref{fig:line1} and \ref{fig:line2} present them separately, together
with the main contributors according to the list of transitions in
Table~\ref{table:1} and Table~\ref{table:2}.
For completeness, we also discuss in this subsection the feature which is
associated to the $\HeIlevel{2}{1}{S}{0}\rightarrow\HeIlevel{2}{1}{P}{1}$
singlet-singlet transition. Figure~\ref{fig:line3} presents the contribution of
this line to the total spectrum in that region.  Although this particular line
is negative, the overall spectrum in the region is positive due to the added
contribution of other lines.
%


%
\subsubsection{First negative feature ($\nu \approx 145$~GHz). }
As Fig.~\ref{fig:line1} indicates, the largest negative contribution is coming
from one of the $10830~\AA$ fine-structure lines. The appearance of this feature
in absorption can be understand as follows. The channel connecting the
$\HeIlevel{2}{3}{S}{1}$ triplet level with the singlet ground state via the
two-photon decay is extremely slow ($\sim\pot{4}{-9}$~s$^{-1}$), and therefore
renders this transition a ``bottleneck'' for those electrons recombining through
the $\HeIlevel{2}{3}{S}{1}$ triplet state.
Since the electrons in the $\HeIlevel{2}{3}{P}{1}$ triplet level can reach the
$\HeIlevel{1}{1}{S}{0}$ level via the much faster ($\sim 177$~s$^{-1}$)
\HeIInt-transition, this provides a more viable route.
On the other hand the $\HeIlevel{2}{3}{P}{0}$ and $\HeIlevel{2}{3}{P}{2}$ do not
have a direct path to the singlet ground state. But as one can see from
Table~\ref{table:2} this restriction can be avoided by taking the route
$\HeIlevel{2}{3}{P}{0/2}\rightarrow\HeIlevel{2}{3}{S}{1}\rightarrow\HeIlevel{2}{3}{P}{1}\rightarrow\HeIlevel{1}{1}{S}{0}$.
The relative amplitude of these lines seen in Fig.~\ref{fig:line1} also suggests
this interpretation.

\subsubsection{Second negative feature ($\nu \approx 270$~GHz). }
The second overall negative feature in the bound-bound $\ion{He}{i}$
recombination spectrum is mainly due to the superposition of the negative
$5877~\AA$ and positive $6680~\AA$-lines (see Fig.~\ref{fig:line2}). Here it is
interesting that in Table~\ref{table:2} no strong positive triplet-singlet
transition appear, which actually starts with a
$\HeIlevel{3}{3}{D}{2}$-state. However, as Table~\ref{table:1} shows a strong
flow from the $\HeIlevel{3}{3}{D}{2}$-state to higher $F$-level exists, again
permitting electrons to pass to the singlet-ground level because of
singlet-triplet mixing.
This then also contributes to the close-by emission feature via the chain
$\HeIlevel{2}{3}{P}{1}\rightarrow\HeIlevel{3}{3}{D}{2}\rightarrow\HeIlevel{4}{3}{F}{3}
\rightarrow\HeIlevel{3}{1}{D}{2}\rightarrow\HeIlevel{2}{1}{P}{1}$.
\subsubsection{The spectrum in the vicinity of $\nu \approx 80$~GHz. }
Fig.~\ref{fig:line3} shows that in this spectral region, there is a clear
low-intensity feature in the overall spectrum, which is produced by the
$\HeIlevel{2}{1}{S}{0}\rightarrow\HeIlevel{2}{1}{P}{1}$ singlet-singlet
transition, that contributes as a negative line.
Comparing the $\HeIlevel{2}{1}{S}{0}$ two-photon decay rate
($A_{\HeIlevel{2}{1}{S}{0}-\HeIlevel{1}{1}{S}{0}}$), with the transition rate to
the $\HeIlevel{2}{1}{P}{1}$, shows that at $z\sim 2500$ the latter is a factor
of $\pot{2}{4}$ larger. Therefore, whenever escape in the \HeILya line
substantially controls the rate of helium recombination, the
$\HeIlevel{2}{1}{S}{0}\rightarrow\HeIlevel{2}{1}{P}{1}$ singlet-singlet
transition appears in absorption. As explained above, accounting for the
hydrogen continuum absorption this is the case at all redshifts of importance.

However, the situation is a bit more involved, since several other transitions
contribute to the negative
($\HeIlevel{3}{3}{D}{2}\rightarrow\HeIlevel{4}{1}{F}{3}$,
$\HeIlevel{3}{3}{D}{3}\rightarrow\HeIlevel{4}{1}{F}{3}$,
$\HeIlevel{3}{3}{D}{2}\rightarrow\HeIlevel{4}{3}{F}{3}$ and
$\HeIlevel{3}{3}{D}{3}\rightarrow\HeIlevel{4}{3}{F}{3}$) and positive
($\HeIlevel{4}{1}{F}{3}\rightarrow\HeIlevel{3}{1}{D}{2}$ and
$\HeIlevel{4}{3}{F}{3}\rightarrow\HeIlevel{3}{1}{D}{2}$) centered at $\nu
\approx 85$~GHz. The superposition of these lines then yields an oscillatory
feature between 80 and 90~GHz, which although always positive, still shows the
clear signature from the 20590~$\,\AA$ line.
Here it is important to realize that several triplet-singlet transitions are
involved, allowing triplet atoms to decay further to the singlet ground
state. This emphasises the importance singlet-triplet mixing for the spectrum,
and in particular well-mixed levels like the low $n$F-states and beyond (mixing
angle $\sim 45^\circ$, see Table 11.12 in \citet{Drake-Book}) provide very
attractive routes.

\subsection{The $\ion{He}{ii}$-recombination spectrum}
\label{sec:HeII.spec}
The recombination history of $\ion{He}{iii}$ is the one which is most close to
the Saha-solution \citep[e.g. see Fig. 15 in][]{Switzer2007III}. Therefore the
release of photons occurs during a shorter period than in the case of
$\ion{H}{ii}$ and $\ion{He}{ii}$ recombination.
In comparison to hydrogen the release of $\ion{He}{iii}$ recombination photon
happens at roughly 4 times higher redshift and temperature (roughly $1400$ for
hydrogen as compared with $6000$ for \ion{He}{iii}).

As Fig.~\ref{fig:final} shows, the high frequency feature always appear on the
red wing of the corresponding hydrogen lines. However at low frequencies the
oscillatory feature drop out of phase with the hydrogen lines.
It is also interesting to see that the $\ion{He}{ii}$ and $\ion{He}{i}$
bound-bound spectra show {\it constructive} (at $\nu\geq 10\,$GHz) and also {\it
destructive} ($\nu\sim 2-5\,$GHz) interference.
As mentioned above, this fact strongly increases the probability to observe
these features in the future.

\begin{figure}
  \centering \includegraphics[width=0.95\columnwidth]{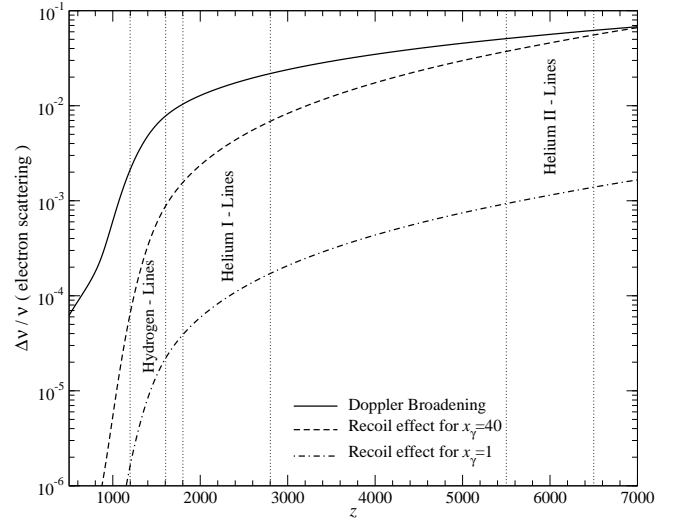}
  \caption{Influence of electron scattering on an initially narrow line for
    different emission redshifts. The vertical lines indicate the
    epochs of recombination at which most of photons are
    emitted. }
  \label{fig:e-scatt}%
\end{figure}

\subsection{Changes of the spectra due to electron scattering}
\label{sec:spec.e-scatt}
The procedure to approximately include the effects of photon scattering off free
electron is outlined in the Appendix \ref{sec:e-scatt}. However, here we neglect
the corrections to the recombination history and recombination spectra arising
from the changes in the escape of photons from the optically thick resonances,
but these are expected to be rather small.

In Fig. \ref{fig:e-scatt} we show the comparison of the Doppler broadening and
recoil term for different redshift. During the epoch of hydrogen recombination
Doppler broadening is less than $1\%$, while it exceeds $\sim 2\%$ during
$\ion{He}{ii}\rightarrow\ion{He}{i}$ recombination, and reaches $\sim 7\%$ at
the beginning of $\ion{He}{iii}\rightarrow\ion{He}{ii}$ recombination.
We also show the strength of the recoil term for $\xg=1$ and $\xg=40$. The
latter case provides an estimate for the equivalent of the Lyman-$\alpha$ line
of the corresponding atomic species.
One can see that during hydrogen recombination the recoil term is completely
negligible. During $\ion{He}{ii}\rightarrow\ion{He}{i}$ recombination the
\HeILya line is shifted by $\lesssim 1\%$ and only for the \HeII Ly-$\alpha$
line the recoil shift is comparable with the broadening due to the Doppler
term. We therefore shall neglect the recoil term for the hydrogen and
$\ion{He}{ii}\rightarrow\ion{He}{i}$ recombination spectrum.
%

\begin{figure}
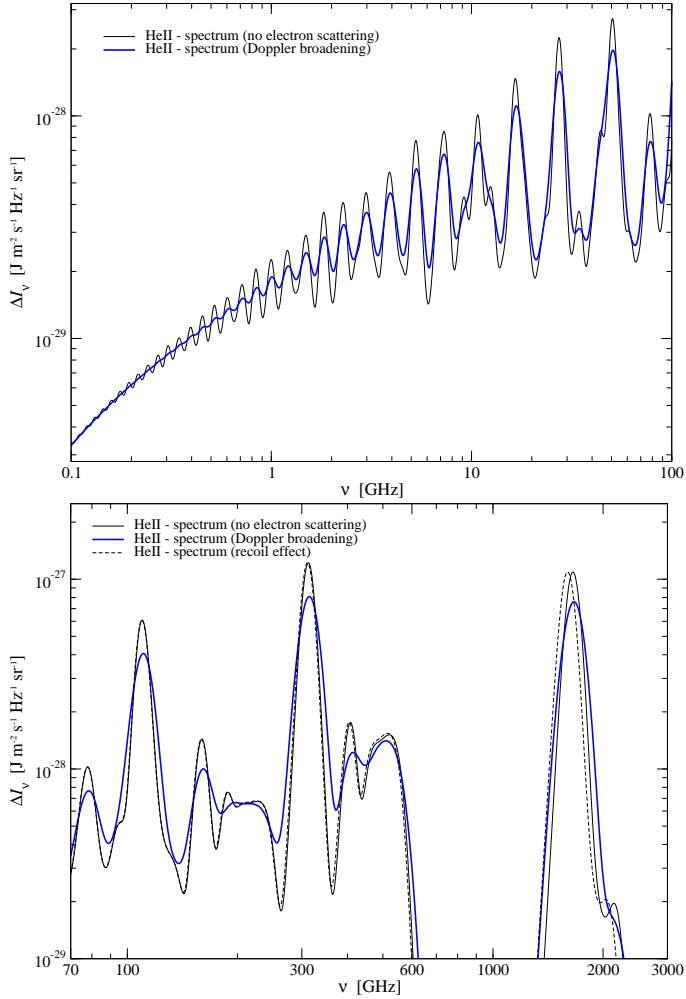

  \centering
  \includegraphics[width=\columnwidth]{eps/DI.e-scatt.low.eps} \\
  \includegraphics[width=\columnwidth]{eps/DI.e-scatt.high.eps}
  \caption{Influence of electron scattering on the
    $\ion{He}{iii}\rightarrow\ion{He}{ii}$ bound-bound recombination
    spectrum. The upper panel shows the changes at low frequencies, where recoil
    is negligible. The lower panel illustrates the importance of the recoil term
    for the \HeII Lyman- and Balmer-series.}
\label{fig:HeIII.e-scatt}%
\end{figure}
Figure \ref{fig:HeIII.e-scatt} shows the importance of the Doppler and recoil
term for the $\ion{He}{iii}\rightarrow\ion{He}{ii}$ bound-bound recombination
spectrum. At low frequencies Doppler broadening strongly lowers the contrast
of the quasi-periodic intensity pattern, while as expected the recoil term is
not important at there. Similarly, the high frequency features are slightly
smoothed out due to the Doppler effect, but the recoil term only becomes
important for the \HeII Lyman- and Balmer-series photons.
However, since in this work we have not yet included the re-processing of
$\ion{He}{ii}$ photons in the continuum of $\ion{He}{i}$ and also the feedback
absorption by hydrogen, we shall not consider the corrections to the
$\ion{He}{iii}$-recombination spectrum due to the recoil term any further. A
more complete treatment of this problem will be left for some future work.
Also, given the overall uncertainty in our model of the neutral helium atom we
did not include the effect of Doppler broadening for the
$\ion{He}{ii}$-recombination spectrum.

%
\section{Discussion}
\label{sec:dis}
In this Section we now critically discuss the results presented in this paper
for the helium recombination spectrum. We expect that an overall $\sim
10\%-30\%$ uncertainty is associated with our modelling of neutral helium, while
neglected physical processes are expected to lead to modification of the
resulting helium bound-bound spectra by $\sim 10\%-20\%$.

We would like to mention that in addition to the aspects discussed below
another $\sim 30-40\%$ rather smooth contribution to the total recombination
emission can be expected from the free-bound components of hydrogen
\citep{Chluba2006b} and helium, possibly with stronger signatures at high
frequencies.

\subsection{Uncertainties in our modelling of the helium atom}

\subsubsection{Completeness of the atomic model}
The probably largest uncertainty is connected with our model of neutral helium.
First of all, for our final bound-bound $\ion{He}{i}$ spectrum (see
Fig.~\ref{fig:final}) we only included levels with $n\leq 30$. As is known from
computation of the hydrogen recombination spectrum \citep{Jose2006,Chluba2006b,
Chluba2007} at low frequencies the level of emission strongly depends on the
completeness of the atomic model. Therefore we expect rather significant
modifications of the $\ion{He}{i}$ spectrum at frequencies below a few
GHz. Computations including up to 100-shells or more are probably necessary. In
the case of the bound-bound $\ion{He}{ii}$ spectrum, for which we already
included $l$-resolved 100-shells, the results are probably converged with
similar accuracy as the one for hydrogen (see \citet{Chluba2007} for
discussion).

We also have not considered quadrupole transitions in our
computations. However, from the typical values of the oscillator strengths
\citep[see e.g.][]{Cann2002}, one would expect that their inclusion should
produce small changes.

\subsubsection{Photoionization cross-sections}
The next large uncertainty is due to the use of re-scaled hydrogenic
approximations for the high-$n$ photoionization cross-sections. We expect
differences at a level of $10\%-20\%$ due to these.
Here in particular the exact frequency dependence of the cross-section may
influence the importance of stimulated recombinations, which become important
for excited levels. Even for $n=5$ there are notable differences, when using
hydrogenic formulae instead of the fits by \citet{Smits1996} and
\citet{Smits1999}. Moreover, we find typical differences of the order of 5-10\%
(and in some lines 20\%) between these fits and the photoionization
cross-sections obtained from the TOPbase database \citep{topbase}.

Furthermore, as shown in Fig. 1 of \citet{Chluba2007} for hydrogen, due to the
strong dependence of the Gaunt-factor on $l$, for large $n$ most of the
recombinations actually go via levels with small $l$.
In particular the S and P states of neutral helium should still have
significant non-hydrogenic contribution for $n>10$, which we did not account
for in our model, again yielding a rather large uncertainty for
$\ion{He}{ii}\rightarrow\ion{He}{i}$ recombination.
Due to the full hydrogenic character for the wave functions of the
$\ion{He}{ii}$ atom, there is no significant uncertainty due to the
cross-sections for $\ion{He}{iii}\rightarrow\ion{He}{ii}$ recombination.

\subsubsection{Energies and transition rates}
In terms of level energies and transition rates, our model of the neutral
helium atom is probably accurate on a level of $1\%-10\%$. The main
uncertainty is connected with the neglect of singlet-triplet mixing for
$n>10$.
As Table~11.12 in \citet{Drake-Book} shows, for $n=10$ the P and D states are
still nearly orthogonal, while the F states are reasonably mixed, and mixing is
practically complete for all the other levels.
However, there are reasons why this may not be of so extreme importance: the
highly excited levels ($n\geq 10$) are already very close to the
continuum. Therefore the route via the continuum leads to a quasi-mixing of
the high levels. In addition, the cascade of electrons to lower levels, where
mixing is fully included, is very fast, such that no significant blocking of
electrons in the higher levels is expected. However, the emission of low
frequency photons probably will be underestimated. Here, a more rigorous
analysis is required.

\subsection{Additional physics missing in our computation}

\subsubsection{$\ion{H}{i}$ continuum opacity}
As discussed in Sect.~\ref{sec:HIopacity}, the hypothesis of {\it complete
redistribution} is not valid for the \HeILya line.
This assumption was usually very good in the context of hydrogen lines
\citep{Grachev1989, RybickiDell94}, in particular due to the presence of a huge
amount of CMB blackbody photons, which allow electrons to pass to higher levels
while they are undergoing a resonant scattering event.
However, it is not the case here, because of the additional continuum opacity
due to small traces of neutral hydrogen during helium recombination.

In order to efficiently compute the escape probability in the \HeILya line, we
make an ansatz about its redshift dependence. This means that we fudge the {\it
no redistribution} solution to the escape probability with a certain factor
which is obtained from a diffusion code which treats in detail the escape
problem.
Although this simplification may introduce errors in the frequency spectrum of
the order of few percent, we consider it acceptable given the uncertainty that
we have in the atomic model and the photoionization cross-sections.

We also made the simplification of assuming the validity of {\it complete
redistribution} for the \HeIInt-line. The exact treatment of the escape of
photons in this line may lead also to differences in the spectrum of the order
of ten percent. In agreement with \citet{Switzer2007III} our more detailed
computations (Chluba et al. 2008, in preparation) show that here electrons
scattering plays an interesting role.

Finally, we stress that in this paper, we only considered the detailed
computation of the deviations of the escape probability with respect to the
Sobolev approximation for the \HeILya transition and \HeIInt-line.  Inclusion of
all the other $\HeIlevel{n}{1}{P}{1}-\HeIlevel{1}{1}{S}{0}$ and spin-forbidden
transitions may lead to corrections of the order of $\sim 10$~\% aswell.

\subsubsection{$\ion{He}{i}$ continuum opacity}
The absorption of $\ion{He}{ii}$ Lyman-$\alpha$ photons by the small fraction
of neutral helium atoms during $\ion{He}{iii}\rightarrow\ion{He}{ii}$
recombination will lead to the appearance of additional $\ion{He}{i}$-photons,
just like in the case of hydrogen (see Sect.~\ref{sec:comp_H_abs.H.spec}).
But since the number of photons emitted in the $\ion{He}{ii}$ Lyman-$\alpha$
line is comparable to the total number of helium nuclei, this will be a
notable change.
Most obviously the $\ion{He}{ii}$ Lyman-$\alpha$ line will nearly disappear.
In addition this will accelerate
$\ion{He}{iii}\rightarrow\ion{He}{ii}$-recombination, bringing it even closer to
the Saha-solution.

\subsubsection{Feedback processes}
As mentioned in Sect.~\ref{sec:Xe}, for
$\ion{He}{ii}\rightarrow\ion{He}{i}$-recombination one of the probably most
important processes that we neglected in our computations so far is {\it
feedback}.
As we have seen in Sect.~\ref{sec:He.spectra} (e.g. Fig.~\ref{fig:lya}), the
total number of photons emitted in the \HeILya transition is comparable with
those coming from the spin-forbidden
$\HeIlevel{2}{3}{P}{1}\rightarrow\HeIlevel{1}{1}{S}{0}$ line.
The former has an energy that is larger by $\Delta\nu/\nu\sim 1\%$. Therefore
one expects the \HeILya photons to interact with the \HeIInt-resonance after a
very short period of redshifting.
The maximum of the \HeILya line appears at $z\sim 2550$ (see
Table~\ref{table:2}), such that the bulk of these photons reach the
spin-forbidden transition at $z_{\rm f}\sim 2520$. At this redshift the optical
depth in the spin-forbidden line is $\lesssim 1$, so that this feedback will not
be complete. Still one should check this process more carefully.

As mentioned above, there is some {\it pure} continuum absorption, far
away from the resonances where resonance scattering can be neglected, which is
not included into our program. This process should also lead to the
re-processing of the remaining \HeILya and \HeIILya photon, such that
practically {\it only} the hydrogen Lyman-$\alpha$ line will survive at the end,
but potentially with interesting traces of the recombination history from
earlier epochs.
Also the feedback due to photons emitted in the \HeI
$\HeIlevel{n}{1}{P}{1}-\HeIlevel{1}{1}{S}{0}$-series (see \citet{Switzer2007I}
and also \citet{Chluba2007b} for more detail) and similarly for
$\ion{He}{ii}$, should lead to some modifications. However, these are expected
to be rather small.
 
\subsubsection{Two-photon decays}
The simplest addition to the two-photon processes is the inclusion of {\it
stimulated emission} as suggested earlier for hydrogen \citep{Chluba2006} and
also included by \citet{Switzer2007II} for helium. These should modify the 2s
two-photon continua at the percent level. However, we have shown that when
accounting for the effect of hydrogen continuum absorption on
$\ion{He}{ii}\rightarrow\ion{He}{i}$ recombination, only $8\%$ of all helium
atoms reach the ground state via this channel. Hence one does not expect large
changes in the $\ion{He}{i}$ recombination spectrum.

For the epoch of $\ion{He}{iii}\rightarrow\ion{He}{ii}$ recombination this may
be a bit different, since electrons in higher levels will feel the change in
the support of the levels from below, because at that time the two-photon
decay channel is more important.
In our computations $\sim 44\%$ of all electrons reach the ground-state of
\HeII via the two-photon channel, and the rest passes trough the \HeII
Lyman-series.
In addition one should include the re-absorption of escaped helium
Lyman-$\alpha$ photons by the two-photon process as discussed by
\citet{Kholupenko2006} for the case of hydrogen.

Also one could think about the two-photon decays from higher levels
\citep{Switzer2007I, Chluba2007c}, but both in terms of additional photons and
increase of the overall rate of recombination one expects corrections at a
level less than 1\%.

\subsubsection{Collisional processes}
In the computations for this paper, collisional processes have not been taken
into account.  As discussed in \cite{Chluba2007} for the case of the hydrogen
recombination spectrum, because of the large entropy of the Universe,
collisional processes only modify the populations of the hydrogen levels for
very high shells. In that paper it is shown that $l$-changing collisions need to
be included only for shells above $n \ga 30-40$, while $n$-changing collisions
can be neglected even for shells as high as $n\approx 100$.

In the case of helium recombination, the same qualitative behaviour is
expected. Although in this case there are more electrons and protons per helium
atom than in the case of hydrogen, we still expect a small effect of collisions,
and which mainly would affect the high-$n$ shells, i.e. it would only have an
impact on the low frequency tail of the recombination spectrum presented in
Figure~\ref{fig:final}. A detailed consideration of the importance of collisions
on the results will be left for a future work.

%
\section{Conclusion}
\label{sec:con}
We have presented detailed computations of the contributions to the
cosmological recombination spectrum due to bound-bound transitions in
primordial helium.
The re-processing of \HeILya and \HeIInt-photons by neutral hydrogen has been
taken into account, yielding a significant acceleration of
$\ion{He}{ii}\rightarrow\ion{He}{i}$ and hence much more narrow features than
without the inclusion of this process. In addition, some hydrogen photons are
released prior to the actual epoch of hydrogen recombination around $z\sim 1100-
1500$, with distinct traces due to the hydrogen Ly-$\alpha$ transition (see
Fig.~\ref{fig:final}).

Probably the most interesting result is the presence of two negative features in
the $\ion{He}{ii}\rightarrow\ion{He}{i}$ recombinational spectrum.  This is
qualitatively different from any of the other spectra discussed so far
($\ion{H}{i}$ and $\ion{He}{ii}$). One of those negative features is associated
to fine-structure transitions in neutral helium.

\begin{figure}
  \centering
  \includegraphics[width=\columnwidth]{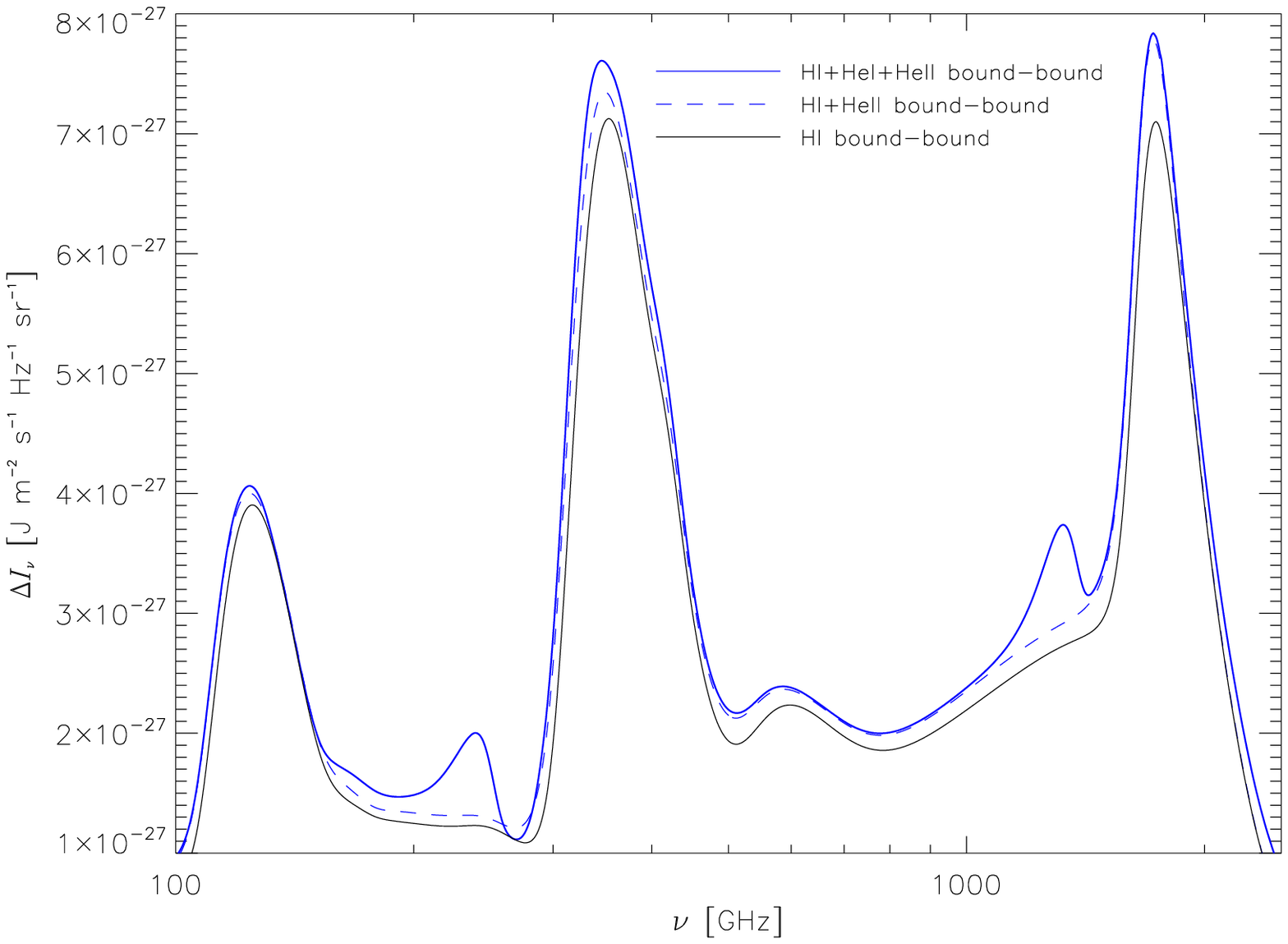} \\
  \includegraphics[width=\columnwidth]{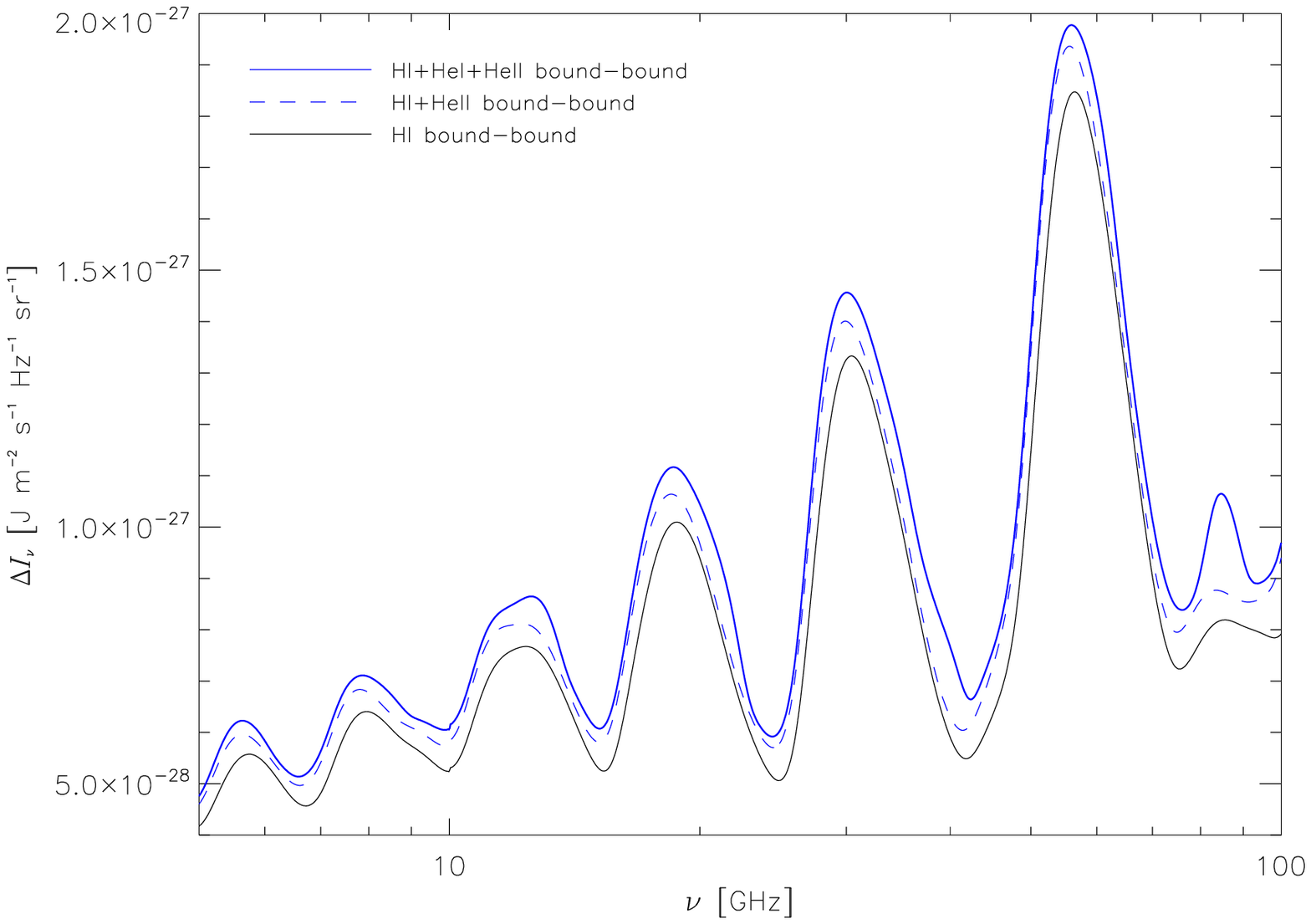}
  \includegraphics[width=\columnwidth]{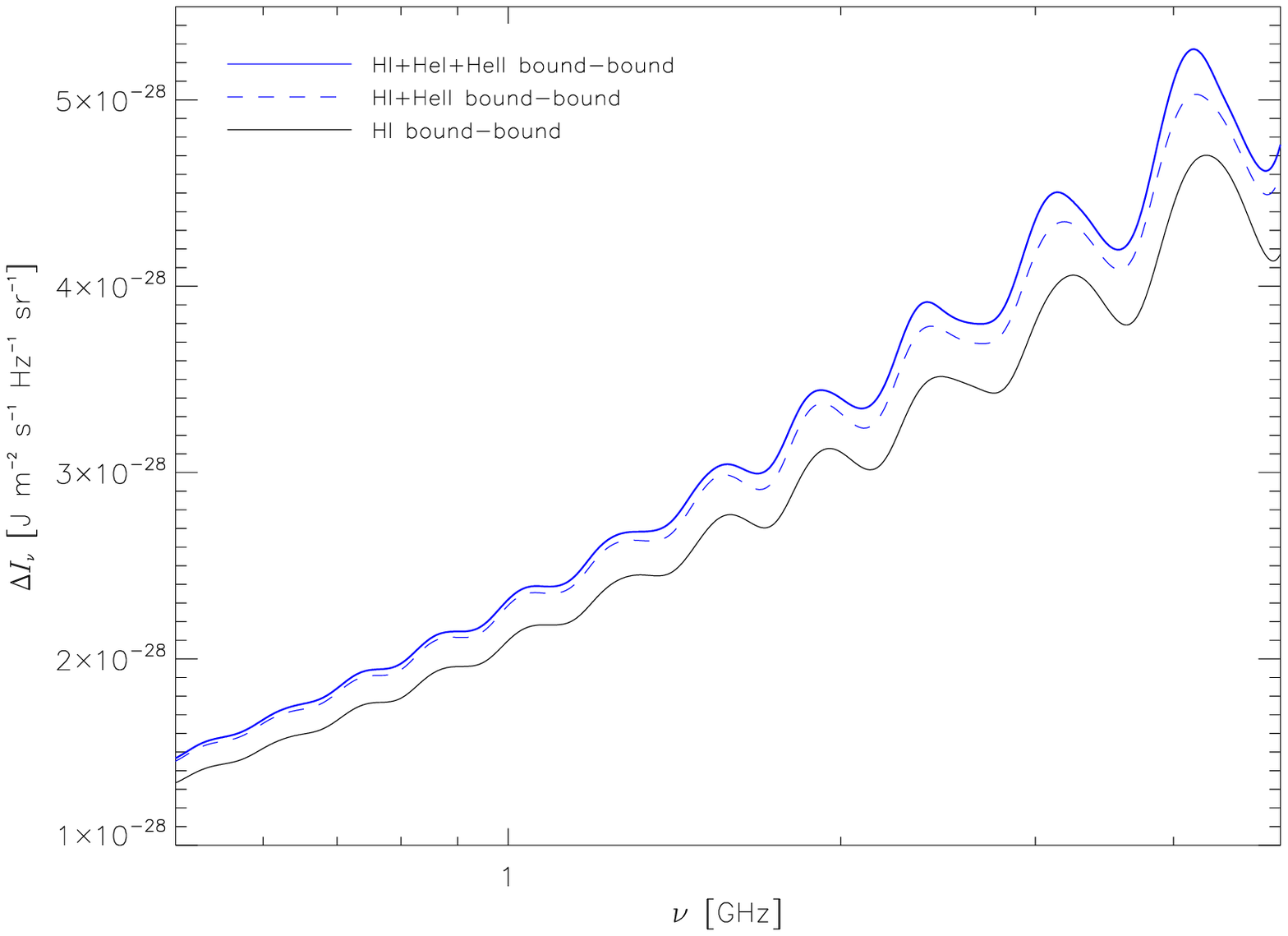}
  \caption{Relative contribution of the \HI, \HeI and \HeII bound-bound
  recombination spectra to the total spectrum at high- (top), intermediate-
  (middle) and low- frequencies (low). Helium recombination spectra (both \HeI
  and \HeII) modify the shapes of the existing hydrogen features, shift the
  peaks positions and introduce new features which represent changes of 30-40\%
  respect to the \HI recombination spectrum alone. }
\label{fig:details}%
\end{figure}

As illustrated in Fig.~\ref{fig:final}, the {\it total} cosmological
recombination spectrum contains non-trivial signatures of all recombination
epochs. We emphasize this fact in Figure~\ref{fig:details}, were we present a
detailed view, using linear intensity scale, of three regions in the
recombination spectrum covering the low, intermediate and high frequency domain.
Although the relative number of helium to hydrogen nuclei is rather small ($\sim
8\%$), constructive and destructive interference of the oscillatory emission
patterns at low frequencies, and strong non-overlapping lines at high
frequencies may provide a unique opportunity to determine some of the key
cosmological parameters, and to confront our current picture of recombination
with experimental evidence.
Interestingly the signatures due to helium may allow a direct determination of
its relative abundance, much before the first appearance of stars, and as
pointed out in \citet{RichnessBeauty}, these measurements do not suffer from
limitations set by cosmic variance.

As we outlined in Sect.~\ref{sec:dis}, several neglected processes have to be
studied in connection with helium recombination, in order to obtain definite
predictions, possibly with additional revisions.
Nevertheless, all the results presented here strongly depend on our
understanding of atomic physics {\it and} the processes in the early Universe.
Currently in particular the data for neutral helium may still not be
sufficient. Here help from atomic physicist is required in order to increase
the availability of more complete {\it accurate} and {\it user-friendly} atomic
data, in particular for the photoionization cross-sections and transition
rates.

All the numerical predictions for the recombinational lines obtained in this
paper, which were used to produce all the figures in this paper, can be
downloaded from \verb!http://www.iac.es/galeria/jalberto/recomb!.

%
\acknowledgement{ The authors thank I.~L.~Beigman and L.~A.~Vainshtein for many
  useful discussions on the physics of neutral helium, and R. Porter for useful
  discussions about the details of the paper by \citet{Bauman2005}.
  We also acknowledge use of the {\sc Cuba}-Library \citep{Hahn2004}. }

\begin{appendix}

\section{Voigt-profile}
\label{app:Voigt}
Evaluations involving the well-known  {\it Voigt}-profile \citep[e.g. see][]{Mihalas1978}:
\bsub
\label{eq:Voigt}
\beal
\varphi(\nu)=\frac{a}{\pi^{3/2}\,\Delta\nu_{\rm D}}\int_{-\infty}^\infty
\frac{e^{-t^2}\id t}{a^2+(x-t)^2}=\frac{\phi(\nu)}{\Delta\nu_{\rm D}},
\end{align}
are usually extremely time-consuming. However, convenient approximations can be
given in the very distant wings and also close to the center of the line.
In Eq. \eqref{eq:Voigt} $x=\frac{\nu-\nu_0}{\Delta\nu_{\rm D}}$ denotes the
dimensionless frequency variable, and the Voigt-parameter and Doppler-width of
the line are defined by
\beal
\label{app:def_DnuD}
a&=\frac{A_{21}}{4\pi \Delta\nu_{\rm D}}
\!\!\!\stackrel{\stackrel{\HeIlevel{2}{1}{P}{1}-\HeIlevel{1}{1}{S}{0}}{\downarrow}}{\approx}
\pot{1.6}{-3} \left[\frac{(1+z)}{2500}\right]^{-1/2}
\\
\label{eq:DnuD}
\frac{\Delta\nu_{\rm D}}{\nu_0}&=\sqrt{\frac{2 k\Te}{\mHe c^2}}
\approx
\pot{1.7}{-5} \left[\frac{(1+z)}{2500}\right]^{1/2},
\end{align}
\esub
respectively. Here $\nu_0$ is transition frequency and $A_{21}$ the Einstein
coefficient for spontaneous emission for the considered resonance.
$\mHe\approx 4 m_{\rm p}$ is the mass of the helium atom.
Note that for the spin-forbidden $\HeIlevel{2}{3}{P}{1}-\HeIlevel{1}{1}{S}{0}$
transition the Voigt-parameter is $\sim 170$ times smaller than for the \HeILya
transition.

For $|x|\leq 30$ we use the approximation based on the Dawson integral up to
sixth order as described in \citet[][Sect.~9.2, p.~279]{Mihalas1978}.
In the distant wings of the line ($|x|\geq 30$) we apply the Taylor expansion
\beal
\label{app:Voigt_appr_wing}
\phi_{\rm wings}
\approx \frac{a}{\pi x^2}\left[1+\frac{3-2a^2}{2 x^2}+\frac{15-20a^2}{4 x^4}+\frac{105(1-2a^2)}{8 x^6}\right].
\end{align}
For the \HeILya and spin-forbidden $\HeIlevel{2}{3}{P}{1}-\HeIlevel{1}{1}{S}{0}$
transition we checked that the Voigt function is represented with relative
accuracy better than $10^{-6}$ in the whole range of frequencies and redshifts.
Using Eq. \ref{app:Voigt_appr_wing}, on the red side of the resonance one can
approximate the integral $\chi=\int_{-\infty}^{x}\phi(x') \id x'$ by:
\beal
\label{eq:Int_Wings_appr}
\chi_{\rm wings}=-\frac{a}{\pi x}\left[1+\frac{3-2a^2}{6 x^2}+\frac{3-4a^2}{4 x^4}+\frac{15(1-2a^2)}{8 x^6}\right].
\end{align}
as long as $x \lesssim -30$. Since $a\sim 10^{-3}$, this shows that the distant
wings only a very small fraction of photons is emitted.
Using the symmetry of the Voigt-profile one finds $\chi(x)=1-\chi(-x)$, such
that Eq. \ref{eq:Int_Wings_appr} is also applicable for $x\gtrsim 30$.

%
\section{Computation of the $\Delta P_{\rm esc}$}
\label{app:pesc}
In this appendix, we focus on some numerical issues which are relevant for the
evaluation of the integral in Eq.~\ref{eq:DPesc}, which gives the escape
probability in the case of complete redistribution of the photons in the
resonance.

\subsection{Analytical approximation of $\Delta P_{\rm esc}$}
\label{sec:anaHIopacity}
For the \HeI $\HeIlevel{n}{1}{P}{1}-\HeIlevel{1}{1}{S}{0}$-series photons
$\tauS\gg 1$ at epochs important for helium recombination. In particular, for
these one can also expect that $\tauL\gg\tauc$ at the relevant
redshifts. Therefore the integrand of Eq.  \eqref{eq:DPesc} will cutoff
exponentially due to the factor $e^{-\tauL}$, while the term $1-e^{-\tauc}$
does not change extremely fast.
For the spin-forbidden transitions this condition is not fulfilled.

Using $\chi(x)$ as variable one can rewrite the integral \eqref{eq:DPesc} as
\begin{align}
\label{eq:DPesc_chi}
\Delta P_{\rm esc}=\int_0^1 \id\chi 
\int_0^{1-\chi} 
\!\!\!\tauS\,e^{-\tau_{\rm S}\Delta\chi'} \Big[1-e^{-\tau_{\rm c}(\chi, \Delta \chi')}\Big] \id \Delta \chi',
\end{align}
with $\Delta\chi'=\chi'-\chi$. The problem now is the computation of
$\tau_{\rm c}(\chi, \Delta \chi')$.  From Eq. \eqref{eq:tau_c_appr} for
$[\nu'-\nu]/\nu\ll 1$ it follows
\begin{align}
\tau_{\rm c}(x, x')\approx\etac(\nu) [\nu'-\nu]=\etac(\nu)\Delta\nu_{\rm
  D}[x'-x],
\end{align}
with $\etac=\frac{c\, N^{\rm H}_{\rm 1s} \sigma^{\rm H}_{\rm 1s}(\nu)}{H
  \,\nu}$. Assuming that $\Delta x=x'-x$ is sufficiently small one may write
$\Delta x \approx \Delta\chi'/\phi(x)$. It is easy to estimate that this
approximation is always very good within the Doppler core, while it is rather
crude in the distant wings.
Inserting this into Eq. \eqref{eq:DPesc_chi} it is possible to carry out the
inner integral analytically, yielding:
\begin{align}
  \label{eq:DPesc_chi_1D}
  \Delta P^{1\rm D}_{\rm esc}\approx\int_0^1 \id\chi
  \left\{1-e^{-\tau_{\rm S}(1-\chi)}-\kappa(\chi)
  \left[1-e^{-[\tauS+\tauct(\chi)](1-\chi)}\right]\right\},
\end{align}
with $\tauct(\chi)=\etac(\nu)\,\DeltanuD/\phi(x)$ and
$\kappa(\chi)=\frac{\tau_{\rm S}}{\tauS+\tauct(\chi)}$, where both $\nu$ and
$x$ are functions of $\chi$.
Numerically this integral is much easier to take than the full 2D-integral
given by Eq. \eqref{eq:DPesc}.  As we will show below this approximation works
very well at low redshift.

\subsection{Numerical evaluation of $\Delta P_{\rm esc}$}
\label{sec:numHIopacity}
To carry out the 2-dimension integral \eqref{eq:DPesc} is a cumbersome task.
We used different integrators from the {\sc Nag}\footnote{See
  http://www.nag.co.uk/numeric/} and {\sc Cuba}\footnote{Download available
  at: http://www.feynarts.de/cuba/}-library and only after several independent
attempts finally reached agreement. It is extremely important to include the
full domain of frequencies, extending the integration to the very distant
wings of the resonance. 
However, due to the extreme differences in the Sobolev escape probabilities of
the \HeI $\HeIlevel{n}{1}{P}{1}-\HeIlevel{1}{1}{S}{0}$-series ($\tauS\sim
10^7$) and intercombination lines ($\tauS\sim 1$) different numerical schemes
are required.
In order to assure convergence of our numerical integrator we made sure that
we can successfully reproduce several limiting cases, for which analytical
approximations can be found. In particular, we reproduced the approximation
given in the previous paragraph.

In order to understand the behaviour of the integrand in Eq.~\eqref{eq:DPesc}
we define the inner integral
\begin{align}
\label{eq:F}
F(x)=\tauS\,\int_x^\infty \phi(x)\,\phi(x') \, e^{-\tau_{\rm L}(x, x')}
\Big[1-e^{-\tau_{\rm c}(x, x')}\Big] \id x',
\end{align}
such that $\Delta P_{\rm esc}=\int_{-\infty}^{\infty} F(x)\id x$. For the \HeI
$\HeIlevel{n}{1}{P}{1}-\HeIlevel{1}{1}{S}{0}$-series one always has $\tauS\gg
1$, such that the exponential factor $e^{-\tau_{\rm L}(x, x')}=e^{-\tauS
\Delta\chi'}$ is dominating the behaviour of the integrand.
  For given $x$ we numerically determined the frequency $x'$ such that
  $e^{-\tau_{\rm L}(x, x')}\leq\epsilon$, typically with $\epsilon\sim
  10^{-16}$. This turned out to be rather time-consuming, but we found that
  even within the Doppler core a sufficient estimate for $x'$ could be
  obtained using the wing expansion of the Voigt-profile, yielding the
  condition
\begin{align}
\label{eq:xp_est}
\chi'-\chi\approx\frac{a}{\pi}\left[\frac{1}{x}-\frac{1}{x'}\right]\leq \frac{\epsilon}{\tauS}.
\end{align}
However, here we typically chose $\epsilon\sim 10^{-25}$, in order to achieve
agreement with the more rigorous treatment. For the spin-forbidden transitions
this simplification is not possible, since none of the exponential factors
really saturate. The full range of frequencies $x\leq x'$ had to be considered
in this case. In practise we never went beyond $10^4$ Doppler width.

\begin{figure}
  \centering 
\resizebox{\hsize}{!}{\includegraphics{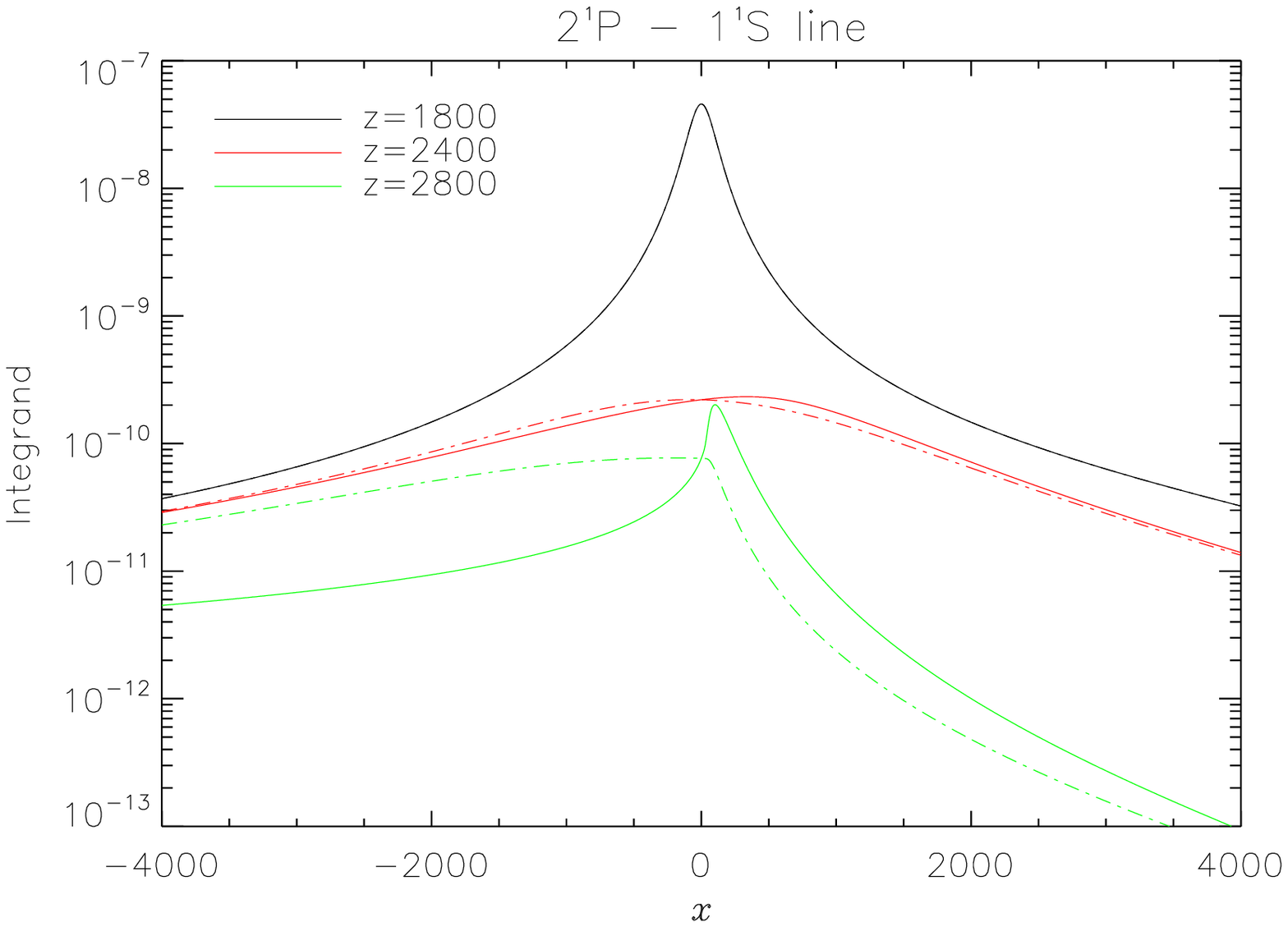}}
\\
\resizebox{\hsize}{!}{\includegraphics{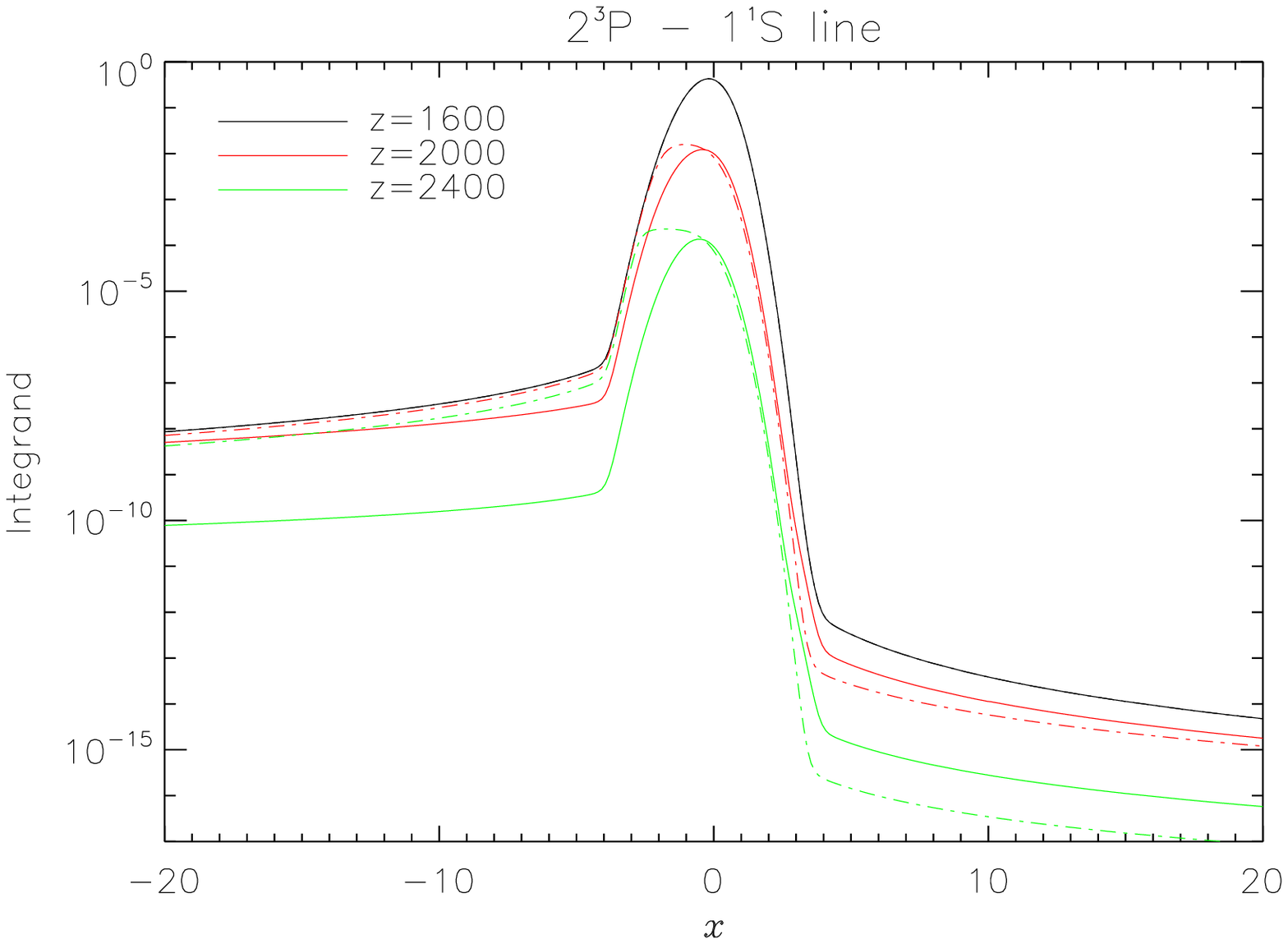}}
\caption{Inner integrand of $\Delta P_{\rm esc}$, as defined by
Eq. \eqref{eq:F}, for the \HeILya line (upper panel) and the spin-forbidden
$\HeIlevel{2}{3}{P}{1}-\HeIlevel{1}{1}{S}{0}$ transition (lower panel).
  For all redshifts, the solid line corresponds to the full numerical integral,
  while the dot-dashed line represents the 1D approximation, as deduced from Eq.
  \eqref{eq:DPesc_chi_1D}.  }
\label{fig:examples.F}
\end{figure}
In Fig.~\ref{fig:examples.F} we show $F(x)$ for the \HeILya line and the
\HeIInt-transition. For the \HeILya line the inner integrand becomes very
broad at high redshifts, with significant contributions to $\Delta P_{\rm
esc}$ out to several thousand Doppler width, while it becomes rather narrow at
low redshifts. However, we found that the outer integral for $\Delta P_{\rm
esc}$ nearly always has to be carried out within a very large range round the
line center. One can also see that the approximation of $F(x)$ following from
Eq.~\eqref{eq:DPesc_chi_1D} works extremely well at low redshifts.
For the \HeIInt-transition the main contributions to $\Delta P_{\rm esc}$ always
come from within the Doppler core and the wing contribution is $\sim
10^{-8}-10^{-7}$ times smaller. For the full numerical integration it therefore
is possible to restrict the outer integral to a few hundred Doppler width.
In Fig.~\ref{fig:examples.F} one can again see that at low redshift the
approximation from Eq. \eqref{eq:DPesc_chi_1D} works very well.

\section{Inclusion of line broadening due to electron scattering}
\label{sec:e-scatt}
The photons released in the process of recombination scatter repeatedly off
moving electrons. In the low temperature limit this process can be described
using the Kompaneets-equation.
Neglecting the small difference in the photons and electron temperature at
redshifts $z\lesssim 500$ and introducing the dimensionless frequency variable
$\xg=h\nu/k\Tg$, neglecting induced effects and the {\it recoil} term for an
initially narrow line, centered at $\xgi$ and released at $\zem$, one can find
the solution \citep{Zeldovich1969, Sunyaev1980}
\beal
\label{eq:DI_nu_Doppler}
\left.\Delta I(\xg, z=0)\right|_{\rm Doppler}=
\frac{x_{\gamma}^3}{x_{\gamma, 0}^3}\,
\frac{\Delta I(\xgi, \zem)}{\sqrt{4\pi\ye}}\times \frac{e^{-\frac{(\ln\xg+3\ye-\ln\xgi)^2}{4\ye}}}{\xgi} ,
\end{align}
where $\Delta I(\xgi, \zem)$ denotes the spectral distortion at frequency
$\xgi$ and redshift $\zem$ without the inclusion of electrons scattering, and
the Compton $y$-parameter is given by
\beal
\label{eq:ye}
\ye(z)=\int_0^z \frac{k\Te}{\me c^2} \frac{c\,\Ne\sigT}{H(z')(1+z')}\id z'
\end{align}
Note that $\Delta I(\xgi, \zem)/x_{\gamma, 0}^3\propto\Delta n(\xgi, z)$,
where $\Delta n$ is the difference of the photon occupation number from a pure
blackbody, is independent of redshift.
As Eq. \eqref{eq:DI_nu_Doppler} show due to the Doppler effect the line
broadens by \citep[compare also][]{Pozdniakov1979}
\beal
\label{eq:Dnu_nu_Doppler}
\left.\frac{\Delta\nu}{\nu}\right|_{\rm Doppler}\sim 2 \sqrt{\ye\,\ln{2}}.
\end{align}
and shifts towards higher frequencies by a factor $e^{3\ye}$.

Also including the recoil term, to our knowledge, no analytic solution to the
Kompaneets equation has been given in the literature. However, to estimate the
effect on the spectrum one can neglect the diffusion term and finds that the
line shifts by 
\beal
\label{eq:Dnu_nu_rec}
\left.\frac{\Delta\nu}{\nu}\right|_{\rm recoil}\sim -\ye\,\xgi 
\end{align}
towards lower frequencies. There is also some line broadening connected with
the recoil effect, but it is completely negligible in comparison with the
Doppler broadening.
From Eq. \eqref{eq:Dnu_nu_rec} is it clear that the high frequency lines will
be affected most.
Note that in contrast to the recoil term Doppler broadening is independent of
the initial photon frequency.

To account for the effect of Doppler broadening on the final spectrum one only
has to integrate Eq. \eqref{eq:DI_nu_Doppler} for fixed $\xg$ over all
possible $\xgi$ for a given transition. The emission redshift $\zem$ of the
contribution can be found with $\nu_0/\nu=1+\zem$. Afterwards in addition the
sum over all transitions has to be carried out, yielding the final results.
To estimate the influence of the recoil effect one can simply add the recoil
shift of each line to the frequency before summing over all possible
transitions.

\end{appendix}

\bibliographystyle{aa} 

\end{document}